\documentclass[fleqn,usenatbib,useAMS]{mnras}

\usepackage[thinc]{esdiff}
\usepackage{multirow}

\usepackage{graphicx}	
\usepackage{amsmath}	
\usepackage{amssymb}	
\usepackage{multicol}        
\usepackage{bm}		
\usepackage{pdflscape}	

\newcommand{\pcm}{cm$^{-3}$}	           
\newcommand{\kms}{km\,s$^{-1}$}            
\newcommand{\msun}{\(\textup{M}_\odot\)}   
\newcommand{\gcm}{g\,cm$^{-3}$}             
\newcommand*\subtxt[1]{_{\textnormal{#1}}}
\DeclareRobustCommand\_{\ifmmode\expandafter\subtxt\else\textunderscore\fi}    

\usepackage[T1]{fontenc}
\usepackage{ae,aecompl}

\usepackage{newtxtext,newtxmath}
\title[Cloud-cloud collisions]{The role of collision speed, cloud density, and turbulence in the formation of young massive clusters via cloud-cloud collisions}
\author[Liow \& Dobbs]{
Kong You Liow$^{1}$\thanks{E-mail: kl457@exeter.ac.uk}
and Clare Dobbs$^{1}$
\\
$^{1}$School of Physics and Astronomy, University of Exeter, Stocker Road, Exeter EX4 4QL, UK
}

\date{Accepted XXX. Received YYY; in original form ZZZ}

\pubyear{2020}

\begin{document}
\label{firstpage}
\pagerange{\pageref{firstpage}--\pageref{lastpage}}
\maketitle

\begin{abstract}
Young massive clusters (YMCs) are recently formed astronomical objects with unusually high star formation rates. We propose the collision of giant molecular clouds (GMCs) as a likely formation mechanism of YMCs, consistent with the YMC \textit{conveyor-belt} formation mode concluded by other authors. We conducted smoothed particle hydrodynamical simulations of cloud-cloud collisions and explored the effect of the clouds' collision speed, initial cloud density, and the level of cloud turbulence on the global star formation rate and the properties of the clusters formed from the collision. We show that greater collision speed, greater initial cloud density and lower turbulence increase the overall star formation rate and produce clusters with greater cluster mass. In general, collisions with relative velocity $\gtrsim 25$ \kms{}, initial cloud density $\gtrsim 250$ \pcm{}, and turbulence of $\approx 2.5$ \kms{} can produce massive clusters with properties resembling the observed Milky Way YMCs. 
\end{abstract}

\begin{keywords}
galaxies: ISM -- ISM: clouds -- stars: formation -- galaxies: star clusters: general
\end{keywords}

\section{Introduction}

Young massive clusters (YMCs) are the most massive recently formed star clusters in galaxies, but their formation is not yet fully understood. The formation of YMCs by large scale gravitational collapse alone appears to disagree with observations of bound clusters \citep{lada_embedded_2003,krumholz_how_2020}, whilst the densities and the small age spreads of YMCs provide extra constraints on how they must have formed. YMCs have cluster ages $\lesssim 100$ Myr, cluster masses $\gtrsim 10^4$ \msun{}, densities $\gtrsim 10^4$ \msun{}/pc$^{3}$, and age spreads of about a few Myr \citep{zwart_young_2010,longmore_formation_2014}. Numerous YMCs have been observed in the Milky Way in the last few decades, including Westerlund 1 \citep{westerlund_1_1961,mengel_westerlund1_2007}, the Arches \citep{nagata_1995,figer_arches_2002} and Quintuplet \citep{quintuplet_1989,figer1999hubble}. However, YMCs are most abundant in interacting or merging galaxies \citep[e.g.][]{mengel_2008,canning_2014,fensch_massive_2019,randriamanakoto_2019}. 

The two most commonly hypothesised formation modes of YMCs are the \textit{in-situ} mode and the \textit{conveyor-belt} mode \citep{longmore_formation_2014}. In the \textit{in-situ} formation mode, YMCs start to form once the precursor giant molecular cloud (GMC) accumulates enough gas mass for star formation, whereas in the \textit{conveyor-belt} mode, YMCs are formed via the concurrent convergence of molecular clouds and any smaller on-going star-forming regions onto a central large star-forming cluster. The latter is most likely to reflect the true YMC formation mode as (i) it is consistent with the hierarchical formation of YMCs observed in the Milky Way \citep{walker_comparing_2016}, (ii) GMCs are unlikely to achieve extremely high densities without significant prior star formation before the \textit{in-situ} formation mode can happen \citep{longmore_formation_2014}, and (iii) even if the gravitational collapse of turbulent GMCs forms clusters, the resultant clusters are often not massive enough or have a longer formation timescale (\cite{dobbs_liow_2020} showed that YMCs cannot be formed solely from gravitational collapse if the initial cloud density is not sufficiently high). A plausible \textit{conveyor-belt} mechanism is cloud-cloud collisions, whereby the collisions can rapidly converge gas onto the collision site and increase the gas mass available for star formation over a short time, triggering star formation \citep{krumholz_how_2020}. The cloud-cloud collision model could explain why YMCs are rare in quiescent galaxies like the Milky Way, where cloud-cloud collisions are both less frequent and less violent \citep{dobbs_frequency_2015}, however common in interacting galaxies, where cloud-cloud collisions occur more frequently \citep{matsui_origin_2012,matsui_galaxy_merger_2019}. In quiescent galaxies, collisions usually happen along the spiral arms where more GMCs are confined \citep{dobbs_gmc_sf_2014}, but mechanisms like supernovae \citep{ntormousi_superbubbles_2011,dawson_supershells_2015,inutsuka_multiple_bubbles_2015} and large-scale turbulence \citep{vazquez_semadeni_turbulence_2012} could induce collisions as well.

Cloud-cloud collisions have been suggested as the formation mechanism for YMCs following observations of highly star-forming regions which are associated with two GMCs with relative velocities from a few \kms{}, up to about 20 \kms{} \citep[e.g.][]{fukui_formation_2015,dewangan_observational_2017,dobashi_cloud_cloud_2019,Enokiya_2019,Sano_2019}. Higher relative velocities are possible too in interacting galaxies like the Antennae galaxies, where collisions with relative velocity $>100$ \kms{} are observed \citep{finn_new_2019,Tsuge_2019}. Often, cloud-cloud collisions are inferred from the unique broad bridge features shown in the gas emission position-velocity diagrams \citep{haworth_isolating_2015}.

Numerical N-body simulations on the formation of YMCs using initial conditions from larger-scale simulations suggest that YMCs can form from hierarchical mergers, whereby the smaller stellar clusters merge and form a massive resultant cluster \citep[e.g.][]{fujii_formation_2015,fujii_initial_2015,grudic_top_2018,li_disruption_2019}. \citet{fujii_formation_2016} followed the formation of YMCs in collapsing GMCs, but they found that these clouds need to be as massive as $\sim 10^5 - 10^6$ \msun{}, as dense as $\sim$ 1000 \pcm{}, and have high velocity dispersions. In the Milky Way, these are the conditions which tend to be associated with regions of cloud-cloud collisions. The hierarchical merger of smaller clusters through pure gravitational interaction also takes longer to achieve a given massive cluster mass without induced external pressure from cloud-cloud collisions, assuming the density is the same for both cases \citep[see Figure 1 of][]{dobbs_liow_2020}.

Alternatively, YMCs may form in more extreme environments.
On larger scales, galactic and cosmological simulations \citep[e.g.][]{renaud_parsec_resolution_2015,dobbs_properties_2017,li_gc_2020,Ma_protoGC_2020} can capture more extreme conditions of clouds interaction to form YMCs, but they often lack the resolution to model the internal physics of the star clusters or systematically test the initial conditions of YMC formation. Simulations of galaxy-galaxy mergers show the formation of massive star clusters \citep[e.g.][]{li_cosmo_I_2017,li_cosmo_II_2018,lahen_merger_2019}. Isolated galaxy simulations by \cite{dobbs_galaxy_2013} showed that cloud mergers can happen along the spiral arms, albeit the resolution was not enough to model cluster formation. \cite{alig_simulating_2018} also simulated a GMC-galactic disc collision and found increased star formation during the collision.

On parsec-scales, hydrodynamical simulations of cloud-cloud collisions \citep[e.g.][]{banerjee_clump_2009,gong_dense_2011,wu_gmc_II_2017} have been performed to model detailed star formation resulting from molecular cloud interactions. Cloud-cloud collisions have been shown effective in forming massive star-forming cores in the shock interface where the clouds meet \citep[e.g.][]{inoue_ccc_2013,takahira_cloud_cloud_2014,takahira_formation_2018,fukui_rapid_2019,sakre_massive_2020}. Many of these emphasise the features of the cores. For example, \cite{matsumoto_2015} studied the mass distribution of the dense cores and found that it follows the classic Salpeter mass function \citep{salpeter_imf_1955}. Colliding cloud simulations by \cite{balfour_star_2015} reproduce star-forming fragmented filaments at the shock interface that are commonly observed in GMCs \citep{hacar_filaments_2013,chira_filaments_2018}. 

Instead of examining the formation of individual stars and star-forming cores at the shock interfaces in this paper, we perform larger-scale cloud-cloud collision simulations which instead focus on the formation of massive clusters. We focus on three parameters, namely the clouds' collision speed, the initial cloud density, and the turbulence in the clouds, and see how they affect the star formation rate and the properties of the massive clusters. Other effects such as magnetic fields and stellar feedback are excluded for simplicity. This paper is structured as follows: in Section \ref{sec:sim}, we describe the details of the simulations and the initial conditions used. In Sections \ref{sec:evolution} and \ref{sec:SFR}, we discuss the effect of the initial conditions on the stellar distribution and the star formation rate. Finally, in Section \ref{sec:properties}, we use a clustering algorithm to identify clusters and investigate the effect of the initial conditions on the properties of the clusters. 

\section{Simulation setup}
\label{sec:sim}
\subsection{Numerical method}

The numerical simulations presented in this paper are performed using \texttt{PHANTOM}, a smoothed particle hydrodynamics (SPH) code for astrophysics \citep{price_phantom_2018}. The smoothing lengths and the densities of the particles are solved simultaneously using the Newton-Raphson method. The $M_4$ cubic spline kernel function \citep{schoenberg_contributions_1946} is used in this SPH calculation. The equations of motion are integrated using the Leapfrog method in `Velocity Verlet' form \citep{Verlet_1967} with individual timesteps for each particle. Gravitational forces are split into short and long-ranged interactions, whereby the former is computed by direct summation over all particles \citep{price_gravity_2007} and the latter is computed by the optimised kd-tree hierarchical grouping of particles \citep{gafton_fast_2011}. Artificial viscosity is introduced in the simulations to handle shocks effectively \citep{monaghan_1997}. We use the standard values for the artificial viscosity parameters of $\alpha\_{min}^{\textnormal{AV}} = 0.1$ and $\alpha\_{max}^{\textnormal{AV}} = 1$ \citep{morris_1997}, but $\beta^{\textnormal{AV}} = 4$ to minimise the effect of particle penetration in high Mach number shocks \citep{price_comparison_2010}. 

Sink particles are introduced to replace the SPH particles when their local densities exceed the critical sink creation density of $\rho\_{sink}=10^{-18}$ \gcm{}. The sink particle accretion radius in our simulations is 0.01 pc. We follow the accretion prescription by \cite{bate_modelling_1995} and more details can be found in Section 2.8 of \cite{price_phantom_2018}. Sink particle mergers are not implemented in our simulations. Note that the sink particles do not represent individual stars. Rather, they simply act as placeholders for high-density SPH particles  and each sink particle represents a few stars. The sink particles interact with all other particles (both sink and gas particles) through gravity, and also the gas particles through mass accretion. We use $N\_{SPH} = 5 \times 10^6$ SPH particles to model each cloud-cloud collision simulation. Taking into account of $\rho\_{sink}$, our chosen resolution can resolve the local Jeans mass for most of our simulations and is sufficient to achieve convergence (See Appendix \ref{ap:restest}).

\subsection{Initial conditions}
\label{ssec:IC}

\begin{table*}
  \begin{tabular}{cccccccc}
    \hline \hline
    $M\_{cloud}$    & $\mathcal{M}\_{turb}$ & $\mathcal{M}\_{coll}$ & $n_0$  & $\rho_0$          & $\mathcal{M}\_{turb}$ at V.E. & $\alpha\_{0,turb}$ & Note \\
    ($10^4$ \msun{}) &                      &                      & (\pcm{}) & ($10^{-22}$ \gcm{}) &                                                &                   &      \\
    \hline \hline
    2.5 & 10 & $0,20,50,100$ & 130 & 5.15 & 11 & 0.41 & Low density, low turbulence       \\
    5   & 10 & $0,20,50,100$ & 256 & 10.3 & 16 & 0.21 & Standard density, low turbulence  \\
    10  & 10 & $0,20,50,100$ & 518 & 20.6 & 22 & 0.10 & High density, low turbulence      \\
    \hline
    2.5 & 20 & $0,20,50,100$ & 130 & 5.15 & 11 & 1.64 & Low density, high turbulence      \\
    5   & 20 & $0,20,50,100$ & 256 & 10.3 & 16 & 0.82 & Standard density, high turbulence \\
    10  & 20 & $0,20,50,100$ & 518 & 20.6 & 22 & 0.41 & High density, high turbulence     \\
    \hline \hline
  \end{tabular}
  \caption{Summary of the initial conditions and key parameters. The mass per cloud $M\_{cloud}$,turbulence $\mathcal{M}\_{turb}$ and collision speed $\mathcal{M}\_{coll}$ are the initial conditions. The consequent parameters are the initial cloud density in \pcm{} $n_0$ and in \gcm{} $\rho_0$, $\mathcal{M}\_{turb}$ at the virial equilibrium, and the ratio of turbulent energy to gravitational potential energy at the start of the simulation $\alpha\_{0,turb}$. The table is separated into low turbulence (top) and high turbulence (lower) simulations.}
\label{tab:ic}
\end{table*}

Initially, two ellipsoidal clouds of mass $M\_{cloud}$ and density $\rho_0$ are created. Each cloud has minor radii of 7 pc and major radius of 16 pc, whereby the major radius is parallel to the $z$-axis. 
We choose ellipsoidal clouds because clouds are observed to be elongated \citep{colomno_gmc_m51_2014, zucker_filaments_2018}. We also see elongated clouds in galactic simulations \citep{duarte_cabral_simulated_clouds_2016}, whereby the filaments in clouds are elongated, and aligned in similar directions, due to shear. Our setup of ellipsoidal clouds also ensures that  clusters form before the gas supply from the two clouds runs out. Each gas clouds contain $0.5 N\_{SPH}=2.5\times10^6$ particles and the gas particles are assigned randomly within the volume of the gas clouds defined. The two clouds are initially set to be 2\% of their major radius apart, as shown in the first panel of Figure \ref{fig:evolve}. The mean molecular weight of the gas is $\mu=2.381$ g/mol and is initially at temperature $T_0=20$ K, which corresponds to a sound speed $c\_{sound}=0.26$ \kms{}. These values of $\mu$ and $T_0$ are the typical values for a giant molecular cloud \citep{dobbs_gmc_sf_2014}. We model our simulations using the isothermal equation of state, following the approximation that the gas in GMCs is typically isothermal \citep{dobbs_gmc_sf_2014,krumholz_star_formation_2015}.

Each cloud collides head-on along the $z$-direction with a collision speed $\mathcal{M}\_{coll}$. The range of collision Mach numbers investigated is $\mathcal{M}\_{coll} = 0, 20, 50$ and 100, where a Mach 0 collision simply means that the clouds are stationary relative to each other. In physical units, the respective relative velocities between the clouds are $\approx 0, 10, 25$ and 50 \kms{}, so our simulated collision speeds covers a wide range of observed realistic relative velocities in potential cloud-cloud collisions \citep[e.g.][]{fukui_formation_2015,dobashi_cloud_cloud_2019,Sano_2019}. For convenience, we name the collision with $\mathcal{M}\_{coll} = 0$ "stationary", $\mathcal{M}\_{coll} = 20$ "low speed", $\mathcal{M}\_{coll} = 50$ "standard speed", and $\mathcal{M}\_{coll} = 100$ "high speed".

The gas clouds are subject to two separate initial supersonic turbulent velocity fields to simulate turbulence, similar to the treatment by \cite{bate_formation_2003}. The initial velocity fields are divergence-free, random and Gaussian. Their power spectrum is $P(k) \propto k^{-4}$, where $k$ is the wavenumber of the field, consistent with Burger's initially-uncorrelated supersonic turbulence \citep{burgers_1948} and Larson's scaling relation \citep{larson_1981}. This power spectrum is slightly steeper than the Kolmogorov's power spectrum of $P(k) \propto k^{-11/3}$ \citep{dubinski_turbulence_1995}. Each velocity field is generated on a $32^3$ grid, then the velocities of the particles are interpolated from the grid. We set the level of turbulence by adjusting the r.m.s. amplitude of the initial velocity fields, such that a greater amplitude is equivalent to stronger turbulence. The levels of turbulence, i.e. the r.m.s. Mach number of the velocity field amplitude used in the simulations are $\mathcal{M}\_{turb} = 10$ ($\approx$ 2.5 \kms{}) and 20 ($\approx$ 5 \kms{}). The range of turbulence chosen simulates the observed velocity dispersion in molecular clouds of similar length scale to ours using Larson's Law \citep[$\approx$ 4 \kms{}, or $\approx$ Mach 16; Figure 1 of][]{larson_1981} and at virial equilibrium (see Table \ref{tab:ic}). Here, we name the turbulence models with $\mathcal{M}\_{turb} = 10$ "low turbulence" and $\mathcal{M}\_{turb} = 20$ "high turbulence". Note that our results are subjected to an uncertainty of $\approx 20$\% due to the choice of random seeds to generate turbulent fields (see Appendix \ref{ap:turbseeds}).

We also change the cloud density by changing the mass per cloud while keeping the shape and the volume of the cloud constant. The range of cloud mass used in this paper is $M = \{2.5,5,10 \} \times 10^4$ \msun{}, which corresponds to an initial cloud density $n_0 = 130$, 256 and 518 \pcm{}. The order of magnitude of $n_0$ is about the same as typical GMC densities \citep{dobbs_gmc_sf_2014}. Again, for convenience, we name the initial cloud density $n_0$ = 130 \pcm{} "low density", $n_0$ = 256 \pcm{} "standard density", and finally $n_0$ = 518 \pcm{} "high density". A summary of the all three initial condition variables and other key parameters is presented in Table \ref{tab:ic}. 

\section{The evolution of clouds and the formation of sink particles}
\label{sec:evolution}

\begin{figure*}
    \centering
    \includegraphics[width=\textwidth,trim={0 12cm 0 0}]{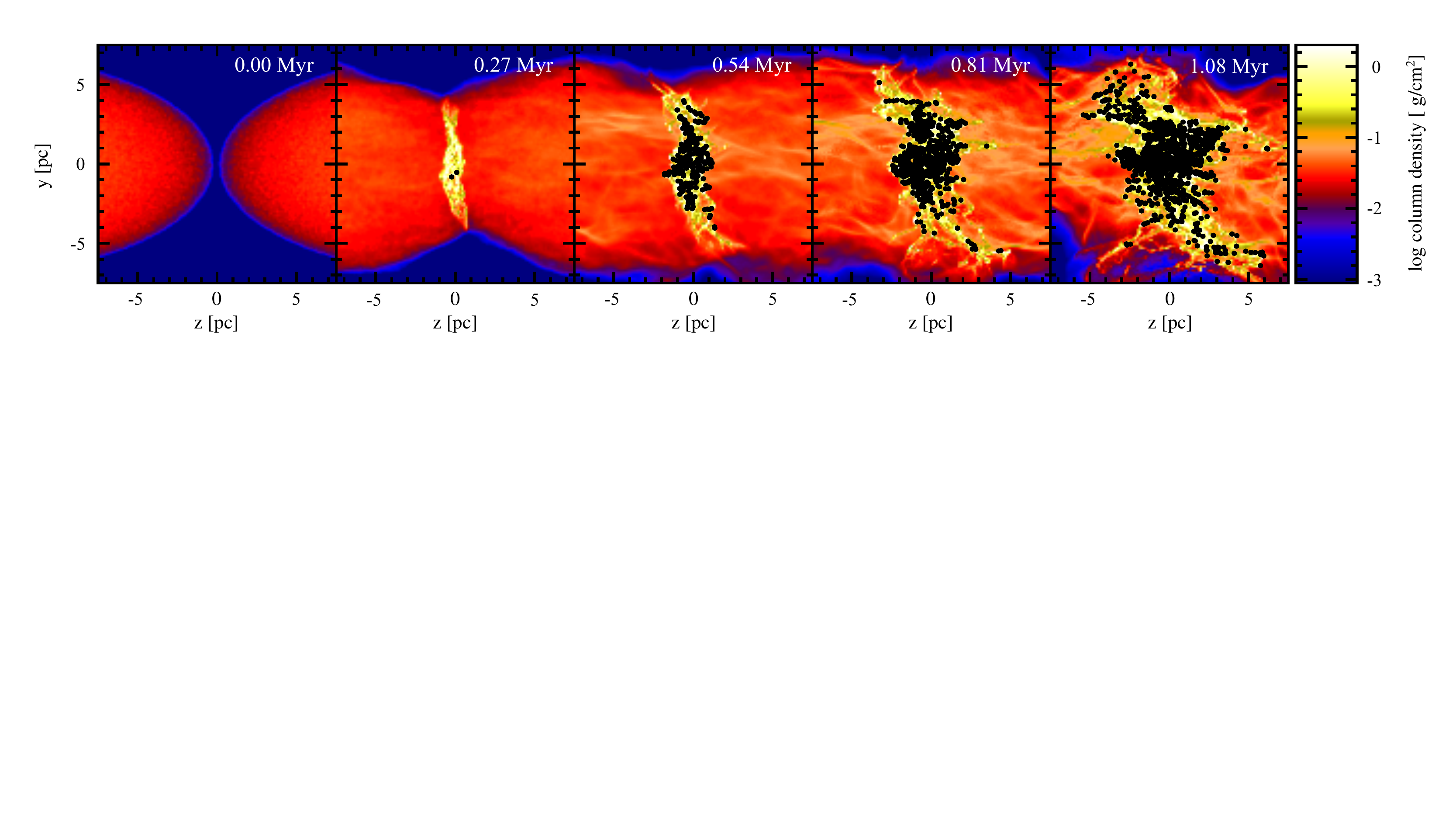}
    \caption{The evolution is shown for the cloud-cloud collision with standard speed, standard density, and low turbulence. The column density plots are as seen perpendicular to the collision axis, taken at different times indicated from the start to $t_{10\%} = 1.08$ Myr. The colour map shows the column density projected on the $(y,z)$-plane. The black dots are the sink particles. }
    \label{fig:evolve}
\end{figure*}

Figure \ref{fig:evolve} shows the evolution of the colliding clouds with standard speed, standard density and low turbulence up to $t_{10\%}=1.08$ Myr, the time when 10\% of the total gas mass is converted into sink particles, following a similar procedure to \cite{balfour_star_2015}. The 10\% efficiency matches the approximate star formation efficiency observed in dense gas clouds \citep{lada_embedded_2003}. Usually, if an isolated turbulent cloud collapses gravitationally, the induced gas flows shock locally forming dense fragments. At points of convergence along the fragments, star formation occurs (or sink particles are inserted \citep{bate_formation_2003}). However, the cloud-cloud collision creates a shock compressed layer parallel to the direction of collision, enabling the shocked layer to fragment, with sink particles forming at the shock interface. This leads to a network of filamentary fragmentation along the $(x,y)$-plane \citep{balfour_star_2015}. For example, the evolution of the clouds shown in Figure \ref{fig:evolve} shows the formation of sink particles at the shock interface, which are induced by the collision. In the last panel of Figure \ref{fig:evolve}, dense filaments are observed away from the shock interface. These filaments are produced via gravitational collapse alone, independently of the collision. Sink particles form in these filaments at a later time. 

The collision and cloud parameters explored in this paper affect the spatial distribution of sink particles at $t_{10\%}$. We first discuss the effect of collision speed and initial cloud density on the spatial distribution using the low turbulence simulations in Section \ref{ssec:lowturb}, then compare the simulations with a different level of turbulence in Section \ref{ssec:highturb}.

\subsection{Low turbulence models}
\label{ssec:lowturb}

\begin{figure*}
    \centering
    \includegraphics[width=\textwidth,trim={0 0 9cm 0}]{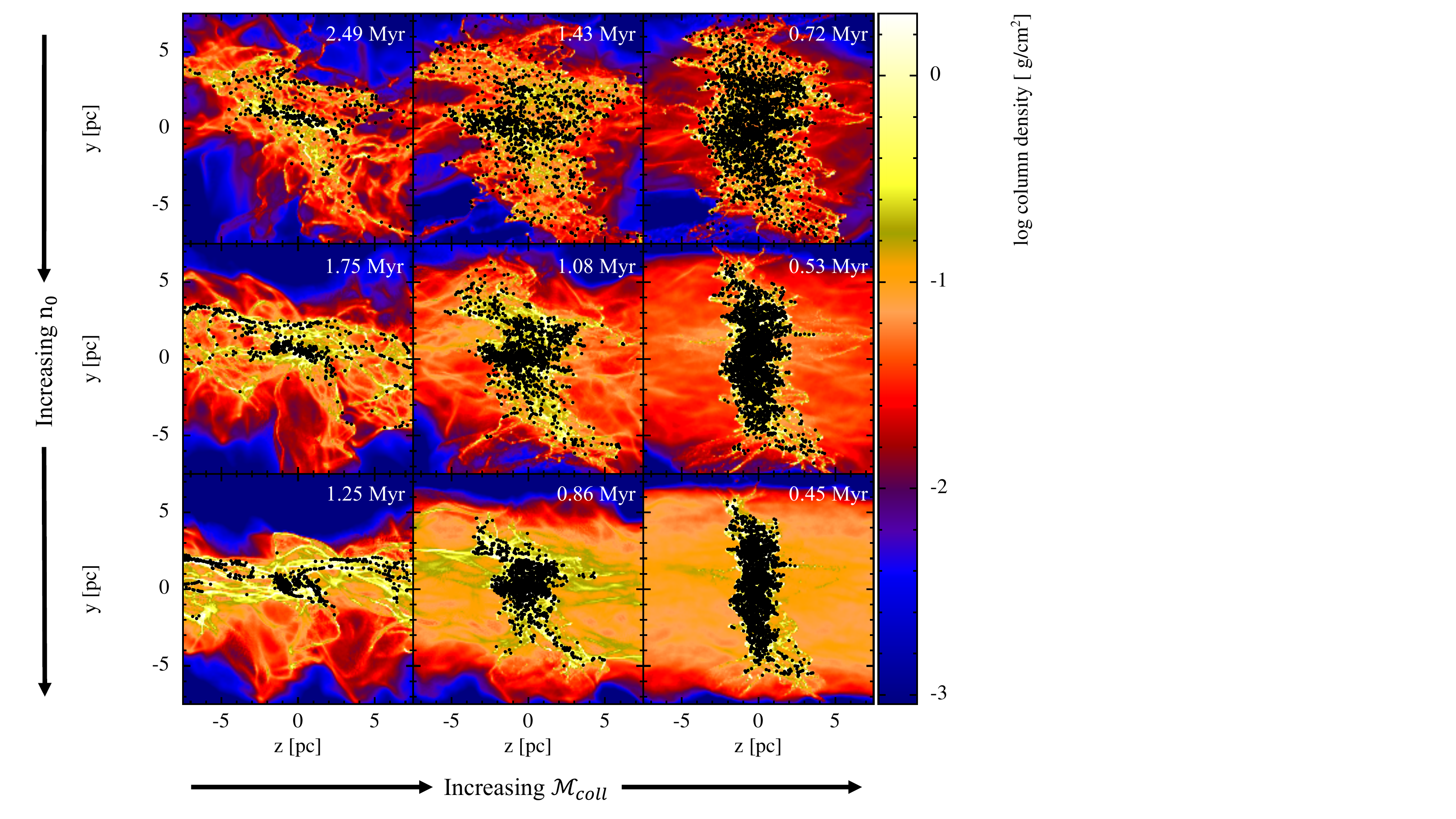}
    \caption{The column density plots for the simulations with Mach 10 turbulence are shown, as seen perpendicular to the collision axis, at $t_{10\%}$ as indicated at the upper-right corner of each subplot. Each subplot shows a box with 7 pc half length from the origin, i.e. the centre of collision. From left to right columns: increasing collision speed ($\mathcal{M}\_{coll} = 20$, 50 and 100). From top to bottom rows: increasing initial cloud density ($n_0 = 130$, 236 and 518 \pcm{}). The colour map shows the column density projected on the $(y,z)$-plane. The black dots are the sink particles.}
    \label{fig:turb1}
\end{figure*}

Figure \ref{fig:turb1} shows the column density plots for all low turbulence simulations at $t_{10\%}$. The rows represent the initial cloud density while the columns represent the cloud collision speed. We limit our view to the 14$\times$14 pc$^2$ box centred at the origin to focus on the sink particles that are formed at the shock interface. The simulations with high density (last row in Figure \ref{fig:turb1}) show clearly the effect of collision speed on the sink particle distribution at the collision site. For the high speed simulation, at $t_{10\%} = 0.45$ Myr, the effect of lateral gravitational collapse along the $(x,y)$-plane is negligible and thus the sink particles are confined in the shock compressed layer, causing their distribution to appear cylindrical. For the standard and low speed cloud-cloud collisions, the sink particles in the shock interface at $t_{10\%}$ have a more spherical distribution. This is because there is more time for gravitational collapse to occur perpendicular to the shock compressed layer. Thus the radial extent of the cluster is reduced along the directions perpendicular to the shock, and the cluster is less elongated. 

For other initial cloud densities (first and second rows in Figure \ref{fig:turb1}), the effect of increasing collision speed on the sink particle distribution is similar. However, we note that the shock interface is more turbulent as the $n_0$ decreases, simply because the ratio of turbulent energy to gravitational potential energy at the start of the simulation $\alpha\_{0,turb}$ is larger for lower density cloud-cloud collisions, i.e. turbulence is more significant than gravity in these models. At the same time, the decrease in $n_0$ reduces the gas mass converged into the shock interface, further reducing its gravitational potential. This causes the sink particle distribution to appear less concentrated and more dispersed.

\begin{figure*}
    \centering
    \includegraphics[width=\textwidth,trim={0 0 6.5cm 0}]{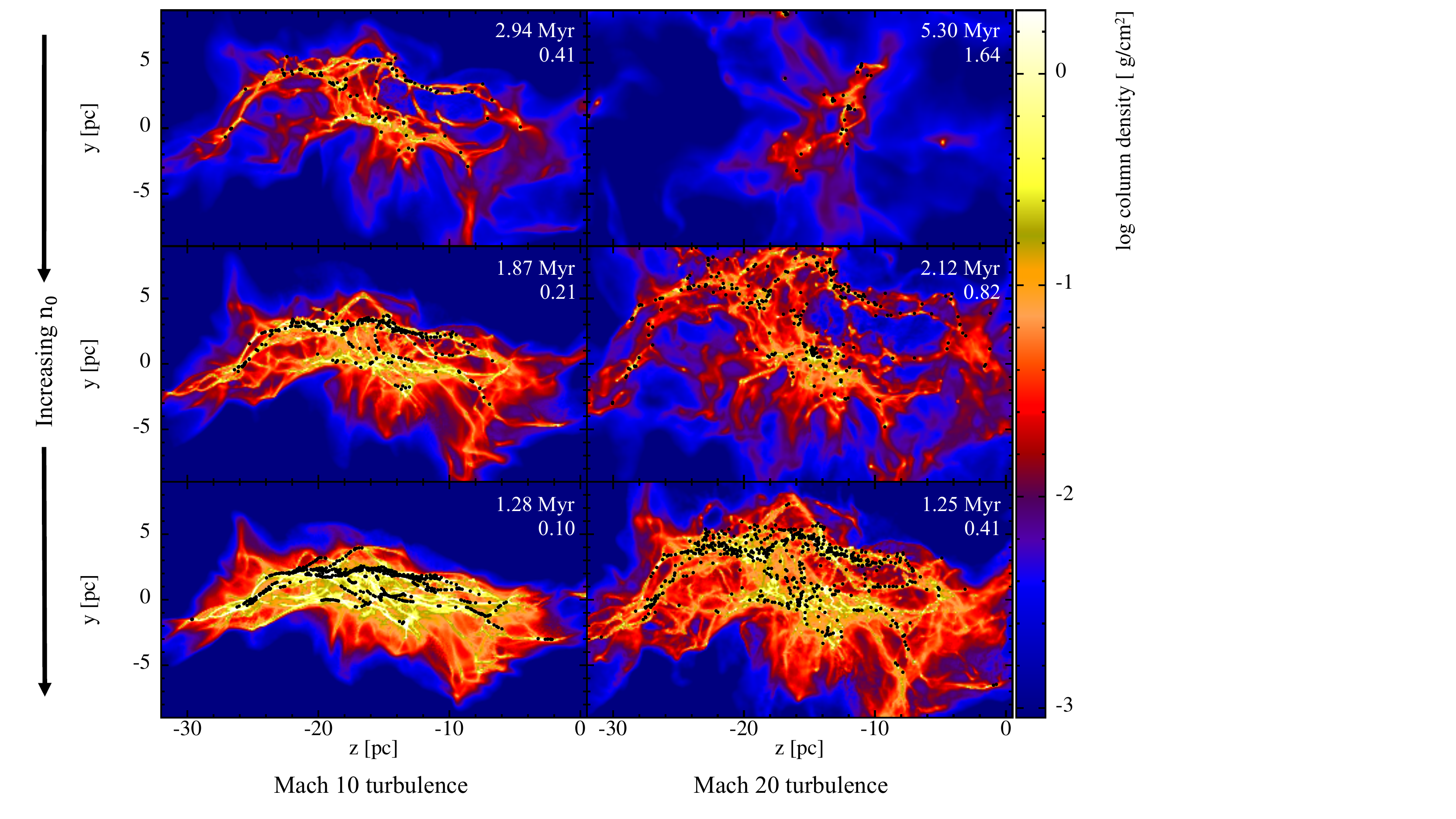}
    \caption{The column density plots for the stationary cases are shown, as seen perpendicular to the cloud major axis, at $t_{10\%}$ as indicated at the upper-right corner of each subplot. Each subplot shows only one of the two evolved clouds, i.e. the cloud that is initially at $z<0$. From left to right columns: increasing turbulence ($\mathcal{M}\_{turb} = 10$ and 20). From top to bottom rows: increasing initial cloud density ($n_0 = 130$, 236 and 518 \pcm{}). The value of $\alpha\_{0,turb}$ for each simulation is shown below the respective $t_{10\%}$. The colour map shows the column density projected on the $(y,z)$-plane. The black dots are the sink particles.}
    \label{fig:stationary}
\end{figure*}

Figure \ref{fig:stationary} shows one of the two clouds at $t_{10\%}$ in the stationary runs. When $\alpha\_{0,turb}$ is large, turbulence dominates and the sink particles are generally far away from each other, whereas with smaller $\alpha\_{0,turb}$, gravity is more dominant. Since the clouds are elongated, gravitational collapse occurs preferentially along the minor axes is and sink particles are formed along the filaments that are approximately parallel to the major axis of the precursor clouds. For both stationary models, the sink particle distributions are approximately filamentary and are not concentrated enough to form massive clusters compared to their colliding clouds counterparts, in agreement with \cite{dobbs_liow_2020} (see Section \ref{sec:properties}).

We see that $t_{10\%}$ is the lowest for the simulations with the highest speed and density, as both conditions result in the greatest gas convergence rate, again in agreement with \cite{dobbs_liow_2020}. For the low speed cloud-cloud collisions, $t_{10\%}$ is not dissimilar to the stationary cases, indicating that the collision is not accelerating the star formation in this case.

\subsection{High turbulence runs}
\label{ssec:highturb}

\begin{figure*}
    \centering
    \includegraphics[width=\textwidth,trim={0 0 9cm 0}]{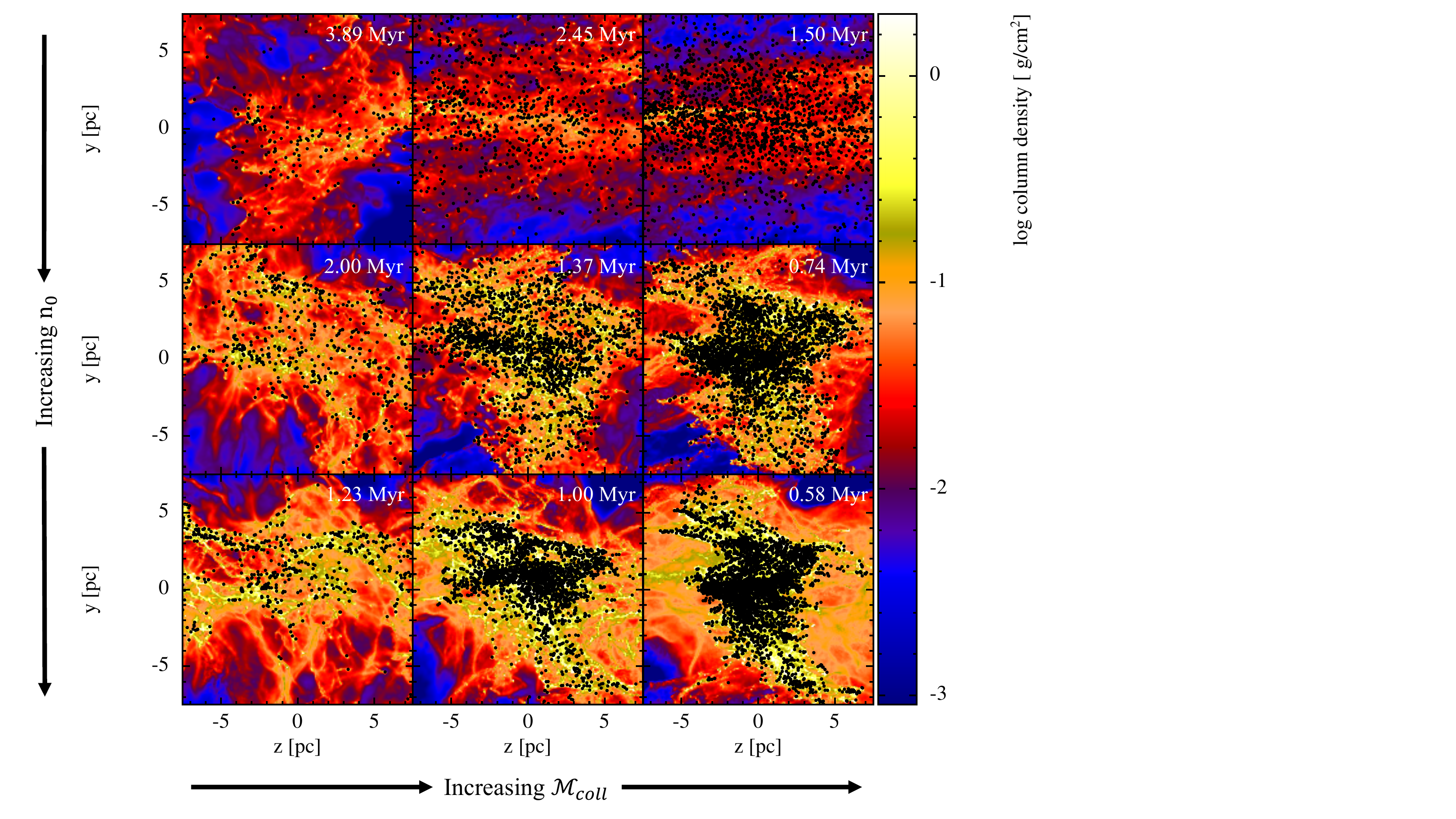}
    \caption{The column density plots for the simulations at Mach 20 turbulence are shown, as seen perpendicular to the collision axis, at $t_{10\%}$ as indicated at the upper-right corner of each subplot. Each subplot shows a box with 7 pc half length from the origin, i.e. the centre of collision. From left to right columns: increasing collision speed ($\mathcal{M}\_{coll} = 20$, 50 and 100). From top to bottom rows: increasing initial cloud density ($n_0 = 130$, 236 and 518 \pcm{}). The colour map shows the column density projected on the $(y,z)$-plane. The black dots are the sink particles.} 
    \label{fig:turb2}
\end{figure*}

Figure \ref{fig:turb2} shows the column density plots for all the high turbulence simulations at $t_{10\%}$. The middle row, which can be compared with the equivalent row for lower turbulence in Figure \ref{fig:turb1}, shows the results for the standard density with high turbulence and different collision speeds. The shock layer is much less apparent in the higher turbulence models, due to the the larger range of velocities of gas colliding, which has the effect of broadening, or washing out the shock. Although higher turbulence induces sink formation at an earlier time, the larger-scale gravitational collapse of the gas is slower (leading to later times for $t_{10\%}$ compared to Figure \ref{fig:turb1}). Similar behaviour was observed by \cite{matsumoto_2015}.

If we compare the higher density models (bottom panels of Figures \ref{fig:turb1} and \ref{fig:turb2}), we see that the effect of turbulence on the sink particle distribution is similar for other collision speeds. For the high turbulence with low density simulations (first row in Figure \ref{fig:turb2}), $\alpha\_{0,turb} > 1$, i.e. the turbulent energy is more than the gravitational potential energy of the precursor clouds. This means that the clouds are already highly turbulent at the beginning of the simulations, and the rate of gas divergence due to turbulence is greater than the rate of gas convergence due to the cloud-cloud collision and global gravitational collapse. There is no clear shock region and the sink particles do not appear to be so evidently located where the clouds collide.

For the other high speed (with standard and high density) cloud-cloud collisions, the sink particles distributions appear to be more three-dimensional compared to their lower turbulence counterparts, albeit not as spherical and compact compared to the standard speed with high density and low turbulence simulation (last row second column in Figure \ref{fig:turb1}). The high compactness of the sink particle distribution in the standard speed model with high density and low turbulence is because the gas convergence rate from all directions are approximately the same, i.e. the free-fall time $\tau\_{ff}$ is approximately the same as the crossing time of the clouds. For the low speed and stationary simulations with high turbulence, the sink particles are again not centrally concentrated enough to form massive clusters. Lastly, we note that for the simulations with high turbulence, $t_{10\%}$ is greater simply because higher turbulence reduces the ability of the gas to converge along the shock, and collapse gravitationally to form sink particles.

\section{Star formation rates}
\label{sec:SFR}
\subsection{General overview}

In our simulations, the total mass of gas particles removed is exactly equal to the total mass of the sink particles formed. The gas particles can be removed either via the formation of new sink particles or by mass accretion onto existing sink particles. We define the star formation rate as 

\begin{equation}
    \dot{M}_* = \diff{M\_{sink}}{t},
    \label{eq:SFR}
\end{equation}

\noindent where $M\_{sink}$ is the total mass of the sink particles. 

\begin{figure*}
    \centering
    \includegraphics[width=\textwidth, trim={2cm 2cm 2cm 0}]{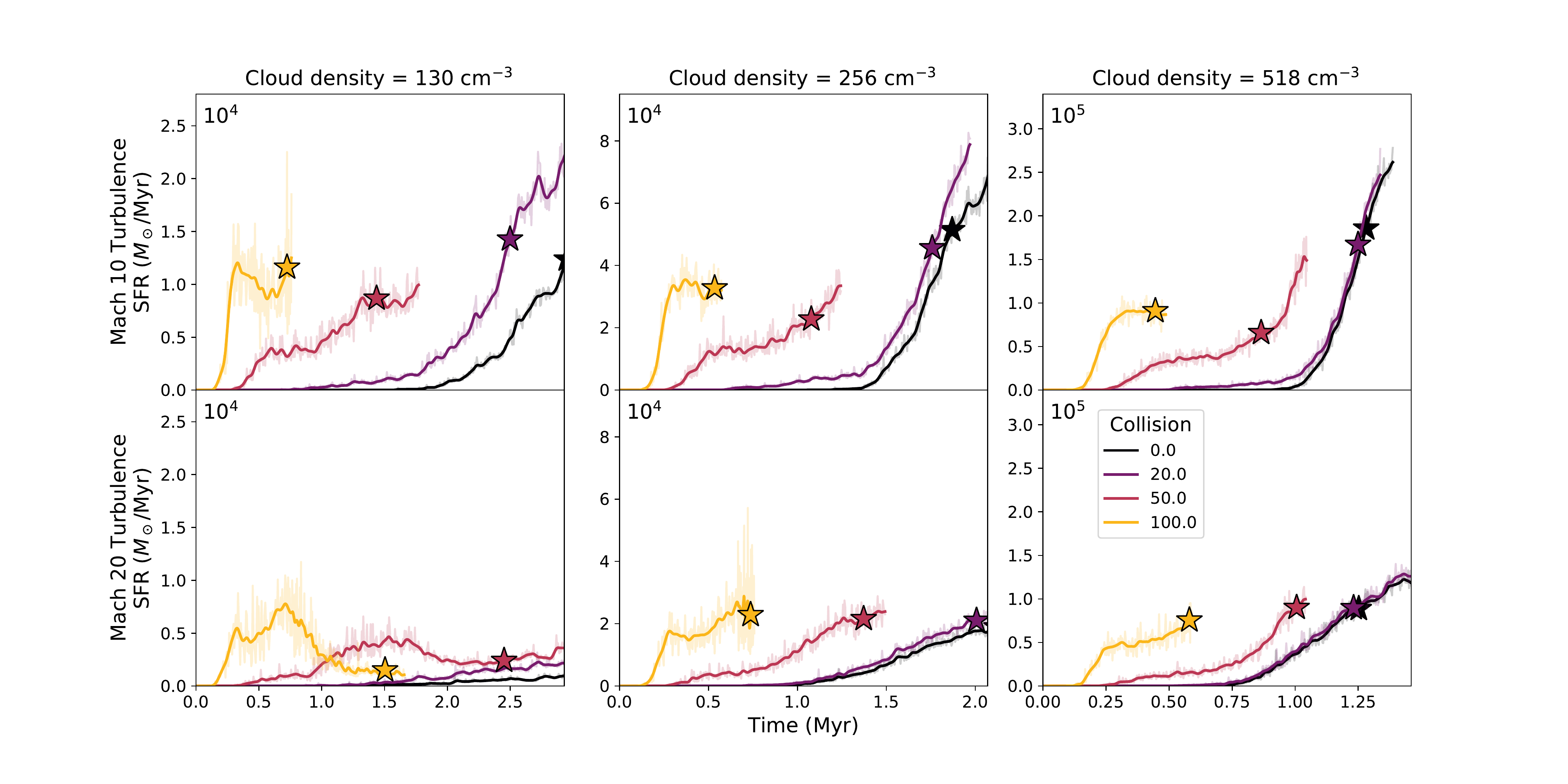}
    \caption{The star formation rates $\dot{M}_*$ are plotted against the absolute time $t$ up to the spherical free-fall time $\tau\_{ff}$ (from left to right columns: $\tau\_{ff} = 2.93, 2.07$ and 1.46 Myr). The collision speed, initial cloud density and level of turbulence are indicated by the colour of the lines, the columns and the rows of the subplots respectively. The translucent lines are the original data, while the solid lines are the 3$\sigma$-Gaussian filtered lines. The star-shaped markers are the star formation rate at $t_{10\%}$. Note that the $y$-axis scales differently for each column.}
    \label{fig:SFR}
\end{figure*}

Equation (\ref{eq:SFR}) is calculated numerically using the forward finite difference method and $\dot{M}_*$ is plotted against $t$ in Figure \ref{fig:SFR} for all simulations up to their free-fall times $\tau\_{ff}$. A 3$\sigma$-Gaussian filter, where $\sigma$ is the standard deviation of the Gaussian kernel, is used to filter the $\dot{M}_*$ values for clearer presentation \citep{scipy_2020}. In general, the star formation rate increases most rapidly in higher speed collision models. For the lower turbulence models (top row of Figure \ref{fig:SFR}), the star formation rates of the low velocity collision and stationary models exceed the higher velocity collision models, but this happens at relatively later times when star formation has already been ongoing for some time. This continued rise occurs as gravity increasingly dominates the star formation. For the high turbulence models (bottom row of Figure \ref{fig:SFR}), the star formation rates reach similar values during the simulations, but these values are reached earlier with higher collision velocities. 

The behaviour of the star formation rate appears different in the higher collision velocity models compared to the stationary and low velocity models. In the former (the red and yellow lines), there is more of a plateau in the star formation rate, whilst in the latter (the purple and black lines), the star formation rate appears to continue accelerating. We suggest that the first plateaus after the initial steep rise result from the star formation mainly induced by the collision, and indeed in these cases most of the star formation occurs in the shock interface (more in Section \ref{ssec:theory}). By contrast, in the low speed and stationary cases, star formation is spread widely through the clouds rather than just at the shock interface. In the colliding case, gravity may still come to dominate the dynamics and lead to accelerated star formation, but this will also depend on whether there is still gas inflowing -- in our high velocity collisions, the gas supply is expected to be exhausted at $t \approx 0.6$ Myr, so most of the gas in the clouds has already participated in the collision at $t_{10\%}$.

There is little difference between the star formation rates of the low speed and stationary runs (this is especially true for high turbulence runs). For the low velocity cloud-cloud collisions with low turbulence, about $30-50$\% of the star formation is concentrated at the shock interface and as such larger mass clusters can formed compared to stationary clouds (see Section \ref{sec:evolution}). However, the overall efficiency of star formation is no higher than the stationary clouds. Figure \ref{fig:SFR} shows that the collision speed likely has to be at least two times greater than the turbulence, i.e. $\mathcal{M}\_{coll} \gtrsim 2\mathcal{M}\_{turb}$ to enable the formation of 
clusters in a shorter timescale. As we see from Figures \ref{fig:turb1} and \ref{fig:turb2}, we also need such velocities for the gas to preferentially fragment at the shock interface, and lead to a strong concentration of filaments and sink particles where the clouds collide. Our result agrees with \cite{matsumoto_2015}, where they found that the flow speed has to be greater than turbulence for filaments to accumulate at the shock interface.

\subsection{Comparison with theoretical expectations}
\label{ssec:theory}
Here, we determine whether the star formation rates due to the collision that we find in the numerical models agree with simple theoretical expectations. For simplicity, we consider the collision of two anti-parallel gas flows in the absence of self-gravity or turbulence. When the gas flows converge, a shock compressed layer is created along the area of convergence. The mass contained in the shock compressed layer is 

\begin{equation}
    M\_s \sim 2 A\_s \rho\_s v\_s t ,
    \label{eq:ms}
\end{equation}

\noindent where $A\_s$ is the cross-sectional area of the shock compressed layer, $\rho\_s$ is the shock density, $v\_s$ is the shock recoiling velocity, and $t$ is the shock accumulation time (approximately the time of simulation here). For the isothermal equation of state, 

\begin{equation}
    v\_s = \frac{-v + v \sqrt{1 + 4/\mathcal{M}\_{coll}^2}}{2}
    \label{eq:vs}
\end{equation}

\noindent and

\begin{equation}
    \rho\_s = \rho_0 \Bigg(1 + \frac{v}{v\_s} \Bigg),
    \label{eq:rhos}
\end{equation}

\noindent \citep{zeldovich_raizer_1966,bate_1995} where $v = \mathcal{M}\_{coll}c\_{sound}$ is the collision speed. Substituting equations (\ref{eq:vs}) and (\ref{eq:rhos}) into (\ref{eq:ms}), we obtain 

\begin{equation}
    M\_s \sim 2 A\_s \rho_0(v + v\_s)t.
    \label{eq:ms2}
\end{equation}

\noindent At large $\mathcal{M}\_{coll}$, the square-root in Equation (\ref{eq:vs}) becomes unity and $v\_s$ vanishes. A similar expression was used by \cite{dobbs_liow_2020} to determine the mass of stars forming due to collisions. Therefore, from Equation (\ref{eq:ms2}), the star formation rate is 

\begin{equation}
    \dot{M}_* \sim \diff{M\_s}{t} \sim 2 A\_s \epsilon \rho_0 v \propto \epsilon n_0 \mathcal{M}\_{coll},
    \label{eq:SFRprop}
\end{equation}

\noindent assuming that $A\_s$ is approximately constant. In reality, not all the gas in shock layer is converted into sink particles, or accreted onto sink particles. Hence, we introduce a conversion efficiency $\epsilon$ in Equation (\ref{eq:SFRprop}).

\begin{figure}
    \centering
    \includegraphics[width=\columnwidth,trim={0 1cm 0 1cm}]{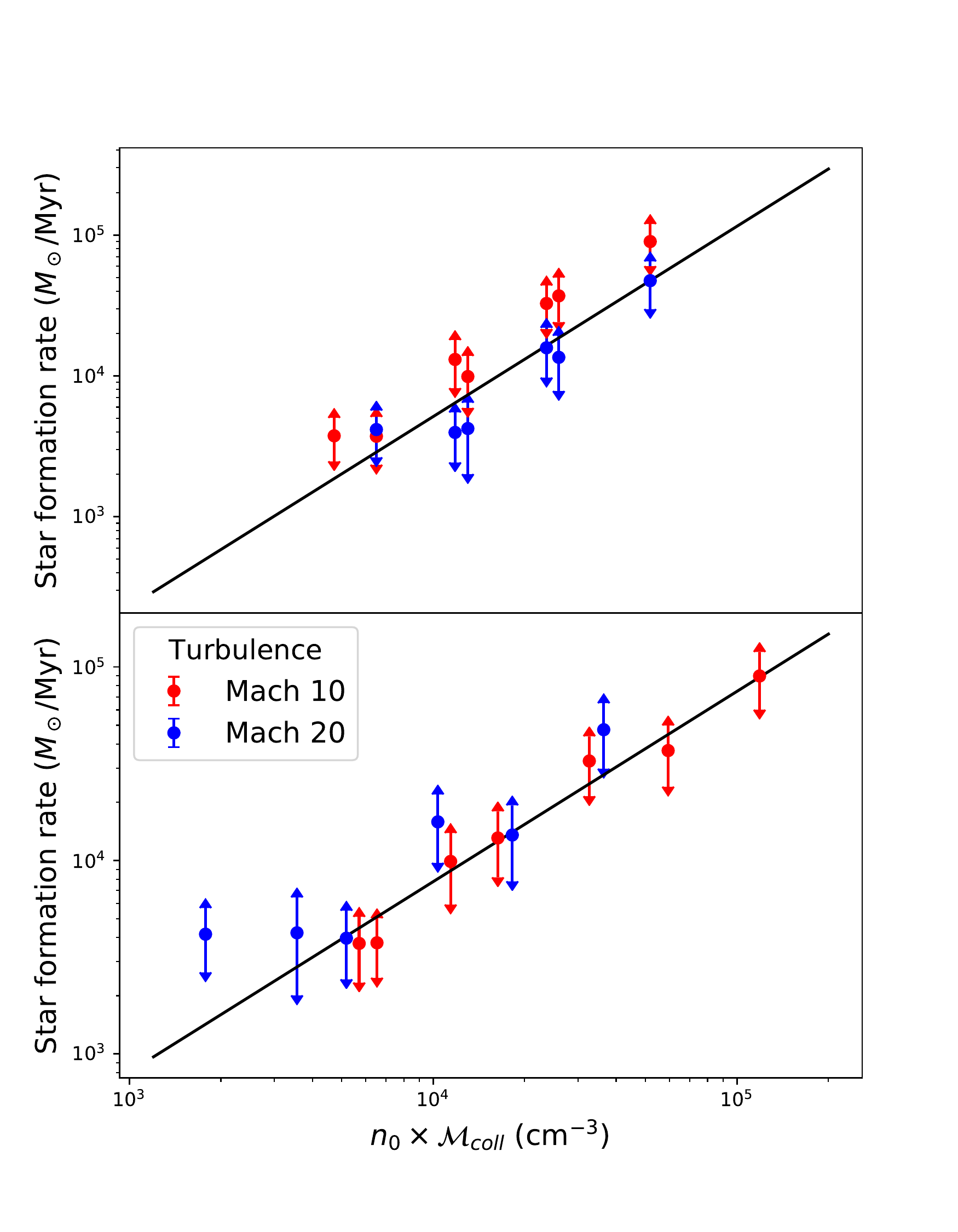}
    \caption{The star formation rates in the collision-dominated regime are plotted against the product of $n_0$ and $\mathcal{M}\_{coll}$. The red points are values from low turbulence models, while the blue points are the values from high turbulence models. In the upper panel (constant $\epsilon$), the black line is the best fit line $\dot{M}_* \propto (n_0 \mathcal{M}\_{coll})^{1.3\pm0.2}$. In the bottom panel ($\epsilon \propto \alpha\_{0,turb}^{-0.68} \mathcal{M}\_{turb}^{-0.32}$), the black fit line is $\dot{M}_* \propto (n_0 \mathcal{M}\_{coll})^{0.99\pm0.09}$.}
    \label{fig:SFRprop}
\end{figure}

To examine Equation (\ref{eq:SFRprop}), we plot the star formation rates for the colliding runs against the product of $n_0$ and $\mathcal{M}\_{coll}$ in the upper panel of Figure \ref{fig:SFRprop}, assuming first a constant $\epsilon$. These star formation rate values are the averages of the first plateaus shown after the initial steep rise typically seen in Figure \ref{fig:SFR}, as during these plateaus the star formation is driven mainly by the collision. We calculate the errors by considering the deviation from the mean and the dispersion caused by the random nature of the star formation in our turbulent collisions (see Appendix \ref{ap:turbseeds}).
We exclude most of the low speed simulations as in those cases the plateaus are not obvious and the star formation rates are similar to the stationary cases. The result shows a positive correlation between the star formation rate and the product $n_0 \mathcal{M}\_{coll}$. The line $\dot{M}_* \propto (n_0 \mathcal{M}\_{coll})^{1.3\pm0.2}$ fits the data, whereby the exponent found overestimates unity (Equation (\ref{eq:SFRprop})) by about 1.5 standard deviations. Similar exponent values are obtained if the star formation rates are plotted against $n_0$ while keeping $\mathcal{M}_{coll}$ constant, and vice versa. 

We also distinguish the star formation rate between the two different levels of turbulence and observe that models with lower turbulence have higher star formation rate at the shock interface. To account for the effect of turbulence in our theory, we incorporate turbulence and the virial ratio at the shock interface as properties affecting $\epsilon$. A simple approach is to introduce a power-law dependence on both the virial ratio $\alpha\_{vir}$ and turbulence $\mathcal{M}\_{turb}$, similar to the power-law fit in \cite{krumholz_turbulent_2005} by considering the lognormal probability distribution function of density in turbulent gas \citep{padoan_fragmentation_2002}, i.e. 

\begin{equation}
    \epsilon \propto \big( \alpha\_{vir}\big)^{-p} \big( \mathcal{M}\_{turb} \big)^{-q}.
    \label{eq:epsilonkrumholz}
\end{equation}

In \cite{krumholz_turbulent_2005}, Equation (\ref{eq:epsilonkrumholz}) is used with $p=0.68$ and $q=0.32$ to fit the star formation rate per free-fall time for a wide range of turbulent-regulated star formation models. To incorporate this into our models, we assume that the relative differences in virial parameters between the models are similar, i.e. those models with high $\alpha\_{0,turb}$ have relatively high virial ratio at the shock interface at $t_{10\%}$, and vice versa. We find that this is roughly true from determining the ratio of kinetic energy to gravitational potential energy in the shocked region. Hence, we let $\alpha\_{vir} \propto \alpha\_{0,turb}$. The use of Equation (\ref{eq:epsilonkrumholz}) is physically intuitive as the star formation rate would be very high for uniform flows, i.e. high density and low turbulence simulations when both $\alpha\_{0,turb}$ and $\mathcal{M}\_{turb}$ are low and hence $\epsilon$ is greater. On the other hand, for high turbulence and low density simulations, the star formation rate is low as the effective filling factor of the dense gas is lower and dense regions would be less likely to collide with each other. Both $\alpha\_{0,turb}$ and $\mathcal{M}\_{turb}$ are greater and hence $\epsilon$ is lower. 

We insert Equation (\ref{eq:epsilonkrumholz}) into Equation (\ref{eq:SFRprop}) with $p=0.68$ and $q=0.32$, and plot the star formation rate against the product of $n_0$ and $\mathcal{M}\_{coll}$ as shown in the bottom panel of Figure \ref{fig:SFRprop}. The exponent is now $0.99 \pm 0.09$, closer to the expected value of unity compared to taking a constant value of $\epsilon$. 
Adjusting the values of $p$ and $q$ by 50\% gives an exponent value of 0.8 -- 1.2. We perform $\chi^2$ goodness-of-fit tests to compare our data with the expected values from the both the regression lines in Figure \ref{fig:SFRprop} and a power law of 1, and also F-tests to compare the regression models and the power law of 1. The result shows that while a constant $\epsilon$ (upper panel of Figure \ref{fig:SFRprop}) is good enough, the regression line that includes the power-law $\epsilon$ (lower panel of Figure \ref{fig:SFRprop}) is better with a 95\% confidence level and is indistinguishable from a power-law of 1, satisfying Equation (\ref{eq:SFRprop}). Thus, including turbulence and the virial ratio according to simple fits by \cite{krumholz_turbulent_2005} provides a way of incorporating these processes into our analytic estimates of the star formation rate and improving on simply assuming uniform flows. Note that in the colliding models, collision-induced star formation occurs before $\tau\_{ff}$ and along the shock front, so the global gravitational collapse of the clouds does not contribute significantly to the star formation rates shown in Figure \ref{fig:SFRprop}. The large-scale gravitational collapse is probably more significant at later times, where the cluster density is high and star formation is still occuring.

\section{The properties of clusters formed via cloud-cloud collisions}
\label{sec:properties}

We use the Density-Based Spatial Clustering of Applications with Noise \citep[DBSCAN,][]{ester_density_based_1996} algorithm to identify the clusters from the 3-dimensional distribution of sink particles.
DBSCAN is a clustering technique that groups together points with similar neighbouring densities \citep{joncour_buckner_khalaj_moraux_motte_2017}, and it has been widely used in observations to identify clusters of various sizes \citep[e.g.][]{zari_structure_2019,winston_2019,Joncour_2018}. The optimum maximum separation between sink particles to be clustered together $\varepsilon$ is 0.5 pc, about 5\% of our simulation length scale. A larger $\varepsilon$ value is likely to include more noise as members of the massive clusters, whilst a smaller $\varepsilon$ value tends to break up massive clusters into smaller subclusters. We choose an arbitrary value of 10 as the minimum number of sink particles in a cluster.

The properties of the most massive clusters formed in the respective colliding models are shown in Table \ref{tab:clusters}, using the definitions of half-mass radius $r\_{hm}$ and half-mass density $\rho\_{hm}$ from \cite{zwart_young_2010} at $t_{10\%}$. The half-mass radius $r\_{hm}$ is the radius of the sphere from the cluster's centre-of-mass that encloses half of the cluster mass $M\_c$. The half-mass density $\rho\_{hm}$ is then the density of this sphere\footnote{The values of $r\_{hm}$ and $\rho\_{hm}$ become more accurate as the number of sink particles in a cluster increases.}. To distinguish the clusters formed in the shock interface from the others, the $z$-component of the cluster centre-of-mass is included, and we define the shock interface as $|z| \leq 3$ pc. In Table \ref{tab:clusters}, we define the cluster age simply as the age of its oldest member, and lastly the virial ratio $\alpha\_c$ as the ratio of the kinetic energy of the cluster to its gravitational potential energy. The position maps of the clusters identified by DBSCAN in Table \ref{tab:clusters} are shown in Appendix \ref{ap:clusters}.

From Table \ref{tab:clusters}, all low turbulence models create clusters with $M\_c \geq 10^3$ \msun{} at the shock interface ($|z| \leq 3$ pc), as expected from the sink particle distributions shown in Figure \ref{fig:turb1} (Section \ref{ssec:lowturb}). Clusters with $M\_c \gtrsim 5000$ \msun{} are formed in the standard density models with higher collision speeds, and the higher density models. For the high turbulence models, we only see such massive clusters in the highest density simulations. Those models with low density or low speed do not form clusters at all at the collision site, as expected from the sink particle distributions shown in Figure \ref{fig:turb2} (Section \ref{ssec:highturb}). In these models, the most massive clusters are $\sim 10^2$ \msun{}, and they are located beyond the collision site. The only exception is Cluster H7 with $M\_c = 1.20 \times 10^3$ \msun{}, but it is likely to be formed independent of the collision and is located at our defined collision site by coincidence. This is further supported by the fact that the next two highest mass clusters have masses of $1.14 \times 10^3$ and $1.04 \times 10^3$ \msun{}, similar in mass to Cluster H7 but are located away from the collision site ($z$-COM $=-8.38$ and $-$4.51 pc respectively). In some of the colliding models (usually with low speed or high turbulence), smaller clusters with $M\_c \sim 10^2$ \msun{} are scattered beyond the shock interface.

The number of sink particles in the cluster ($N$) is the highest for models with greater speed. Nonetheless, the average mass per sink particles is actually the lowest for these models, as the gas particles are less likely to be gravitationally bound when testing for sink creation, or accretion onto sink particles.
The cluster half-mass radius $r\_{hm}$ is larger in high speed simulations, but the calculation of $r\_{hm}$ assumes a spherical cluster distribution and does not take into account of the morphology of the clusters. Therefore, clusters that have two-dimensional-like distribution, which are common for those that formed under high speed collision and have not experienced strong self-gravity yet, tend to have larger $r\_{hm}$. By considering fragmentation of the shocked layer,  \cite{whitworth_star_formation_2016} found that the first sink particle is predicted to form at a time proportional to $n_0^{-1/2}\mathcal{M}\_{coll}^{-1/2}$. Most clusters shown in Table \ref{tab:clusters} include the first sink particle formed in a given simulation. We find that the ages of the clusters follow the same dependence as found by \cite{whitworth_star_formation_2016}, even though we note that measuring the cluster age from the first sink particle is not necessarily optimal, and we will reconsider this in future work.


Similarly to Table \ref{tab:clusters}, Table \ref{tab:clustersusual} shows the most massive clusters identified by DBSCAN in the stationary runs. Those stationary models with higher density or lower turbulence can form clusters with $M\_c \geq 10^3$ \msun{} through smaller-scale gravitational collapse as $\alpha\_{0,turb}$ is low. Only the high density model with low turbulence forms a cluster greater than 5000 \msun{}, but it is unsurprising given its high initial cloud density. Nevertheless, these clusters are still less massive compared to their colliding counterparts, as stationary models lack the external pressure from collision to concentrate gas mass into a central star-forming region. Smaller clusters with $M\_c \sim 10^2$ \msun{} can be found scattered throughout the clouds. These collision-independent formed clusters are usually formed along the dense filaments created via lateral gravitational collapse of the precursor ellipsoidal clouds. Most of the clusters that are 
formed independent of the collision
are formed much later in the evolution of the clouds and have ages similar to, or smaller than their colliding counterparts.
Because star formation is much more widespread in the stationary clouds, compared to the colliding case, a given cluster is much less likely to contain the oldest sink particles compared to the colliding clouds, where most sink particles are concentrated in a central cluster.
The collision-induced massive clusters are formed earlier, as measured by the oldest sink particle they contain, and accrete more mass over a longer timescale. 

\begin{table*}
    \centering
    \begin{tabular}{c|c|c|c|c|c|c|c|c|c|c|c|c}
    \hline \hline
    Name & $n_0$  & $\mathcal{M}\_{coll}$ & $\mathcal{M}\_{turb}$ & $t_{10\%}$ & $N$  & $z$-COM & $M\_c$          & $r\_{hm}$ & $\rho\_{hm}$      & Age   & $\alpha\_c$  \\
    
        & (\pcm{}) &                      &                      & (Myr)       &      & (pc)    & ($10^3$ \msun{}) & (pc)     & (\msun{}/pc$^{3}$) & (Myr) &           \\
     \hline \hline
    L1 & 130 & 20  & 10 & 2.49 & 191  & -1.59 & 1.71  & 0.64 & 786.8  & 1.82 & 0.32 \\
    L2 & 130 & 50  & 10 & 1.43 & 495  & -0.84 & 2.35  & 1.93 & 38.8   & 1.14 & 0.78 \\
    L3 & 130 & 100 & 10 & 0.72 & 1310 & -0.32 & 2.64  & 2.50 & 20.0   & 0.57 & 2.21 \\
    L4 & 236 & 20  & 10 & 1.75 & 412  & -0.37 & 4.62  & 1.15 & 362.3  & 1.26 & 0.36 \\
    L5 & 236 & 50  & 10 & 1.08 & 1102 & -0.43 & 6.39  & 1.64 & 167.0  & 0.82 & 0.69 \\
    L6 & 236 & 100 & 10 & 0.53 & 2748 & -0.21 & 8.77  & 2.82 & 46.8   & 0.41 & 0.84 \\
    L7 & 518 & 20  & 10 & 1.25 & 605  & -0.34 & 6.61  & 0.64 & 3063.7 & 0.82 & 0.45 \\
    L8 & 518 & 50  & 10 & 0.86 & 2314 & -0.19 & 18.20 & 1.46 & 705.6  & 0.63 & 0.58 \\
    L9 & 518 & 100 & 10 & 0.45 & 3384 & -0.15 & 19.00 & 2.67 & 118.8  & 0.33 & 0.44 \\
     \hline
    H1  & 130 & 20  & 20 & 3.89 & 13   & -4.84 & 0.36  & 0.17 & 9471.4 & 1.83 & 0.60  & \\
    H2  & 130 & 50  & 20 & 2.45 & 58   & -4.93 & 0.53  & 0.60 & 297.4  & 2.14 & 0.25  & \\
    H3  & 130 & 100 & 20 & 1.50 & 105  & -7.34 & 0.33  & 0.69 & 120.6  & 1.34 & 8.16  & \\
    H4  & 236 & 20  & 20 & 2.00 & 31   & 4.42  & 0.56  & 0.10 & 57949.8& 0.80 & 1.24  & \\
    H5  & 236 & 50  & 20 & 1.37 & 503  & -2.81 & 2.24  & 1.45 & 88.3   & 1.10 & 1.45  & \\
    H6  & 236 & 100 & 20 & 0.74 & 1836 & -0.71 & 4.85  & 2.75 & 27.9   & 0.61 & 2.98  & \\
    H7  & 518 & 20  & 20 & 1.23 & 107  & -2.49 & 1.20  & 0.66 & 504.7  & 0.76 & 0.97  & \\
    H8  & 518 & 50  & 20 & 1.00 & 1865 & -0.41 & 11.90 & 2.16 & 140.3  & 0.78 & 0.85  & \\
    H9  & 518 & 100 & 20 & 0.58 & 3848 & -0.19 & 15.50 & 2.81 & 83.3   & 0.45 & 1.24  & \\
    \hline \hline
    \end{tabular}
    \caption{The list of the most massive clusters formed in the colliding models with their precursor clouds' initial conditions $\mathcal{M}\_{coll}, \mathcal{M}\_{turb}$ and $n_0$ in the first three columns. The subsequent columns are the time when 10\% of the gas mass becomes sink particles $t_{10\%}$, the number of sink particles $N$, the $z$-component of the cluster's centre-of-mass, the cluster mass $M\_c$, the half-mass radius $r\_{hm}$, the half-mass density $\rho\_{hm}$, the cluster age, and the virial ratio $\alpha\_c$. The table is separated into two sections according to the level of turbulence, which is also indicated in the cluster name (`L': low turbulence; `H': high turbulence).}
    \label{tab:clusters}
\end{table*}

\begin{table*}
    \centering
    \begin{tabular}{c|c|c|c|c|c|c|c|c|c|c|c|c}
    \hline \hline
    Name & $n_0$  & $\mathcal{M}\_{coll}$ & $\mathcal{M}\_{turb}$ & $t_{10\%}$ & $N$  & $z$-COM & $M\_c$          & $r\_{hm}$ & $\rho\_{hm}$      & Age   & $\alpha_c$ \\
    
        & (\pcm{}) &                      &                      & (Myr)       &      & (pc)    & ($10^3$ \msun{}) & (pc)     & (\msun{}/pc$^{3}$) & (Myr) &            \\
     \hline \hline
    L10 & 130 & 0   & 10 & 2.94 & 16   & 15.9  & 0.37  & 0.037 & 847285.9 & 0.76 & 0.34  \\
    L11 & 236 & 0   & 10 & 1.87 & 232  & -15.2 & 2.39  & 1.49    & 87.0   & 0.48 & 0.50   \\
    L12 & 518 & 0   & 10 & 1.28 & 797  & -16.8 & 7.81  & 2.63    & 51.6   & 0.38 & 0.85    \\
     \hline
    H10   & 130 & 0   & 20 & 5.30 & 11    & 14.3     & 0.66     & 0.20    & 10597.5      & 3.26    & 0.27      \\
    H11   & 236 & 0   & 20 & 2.12 & 38    & 15.6     & 0.80     & 0.13    & 46383.3      & 0.92    & 0.35       \\
    H12  & 518 & 0   & 20 & 1.25 & 91   & -11.8 & 1.06  & 1.10 & 93.8   & 0.43 & 1.27    \\
    \hline \hline
    \end{tabular}
    \caption{The list of the most massive clusters formed in the stationary runs with their precursor clouds' initial conditions $\mathcal{M}\_{coll}, \mathcal{M}\_{turb}$ and $n_0$ in the first three columns. The subsequent columns are the time when 10\% of the gas mass becomes sink particles $t_{10\%}$, the number of sink particles $N$, the $z$-component of the cluster's centre-of-mass, the cluster mass $M\_c$, the half-mass radius $r\_{hm}$, the half-mass density $\rho\_{hm}$, the cluster age, and the virial ratio $\alpha\_c$. The table is separated into two sections according to the level of turbulence, which is also indicated in the cluster name (`L': low turbulence; `H': high turbulence). }
    \label{tab:clustersusual}
\end{table*}

\begin{figure}
    \centering
    \includegraphics[width=\columnwidth,trim={0 0 20cm 0}]{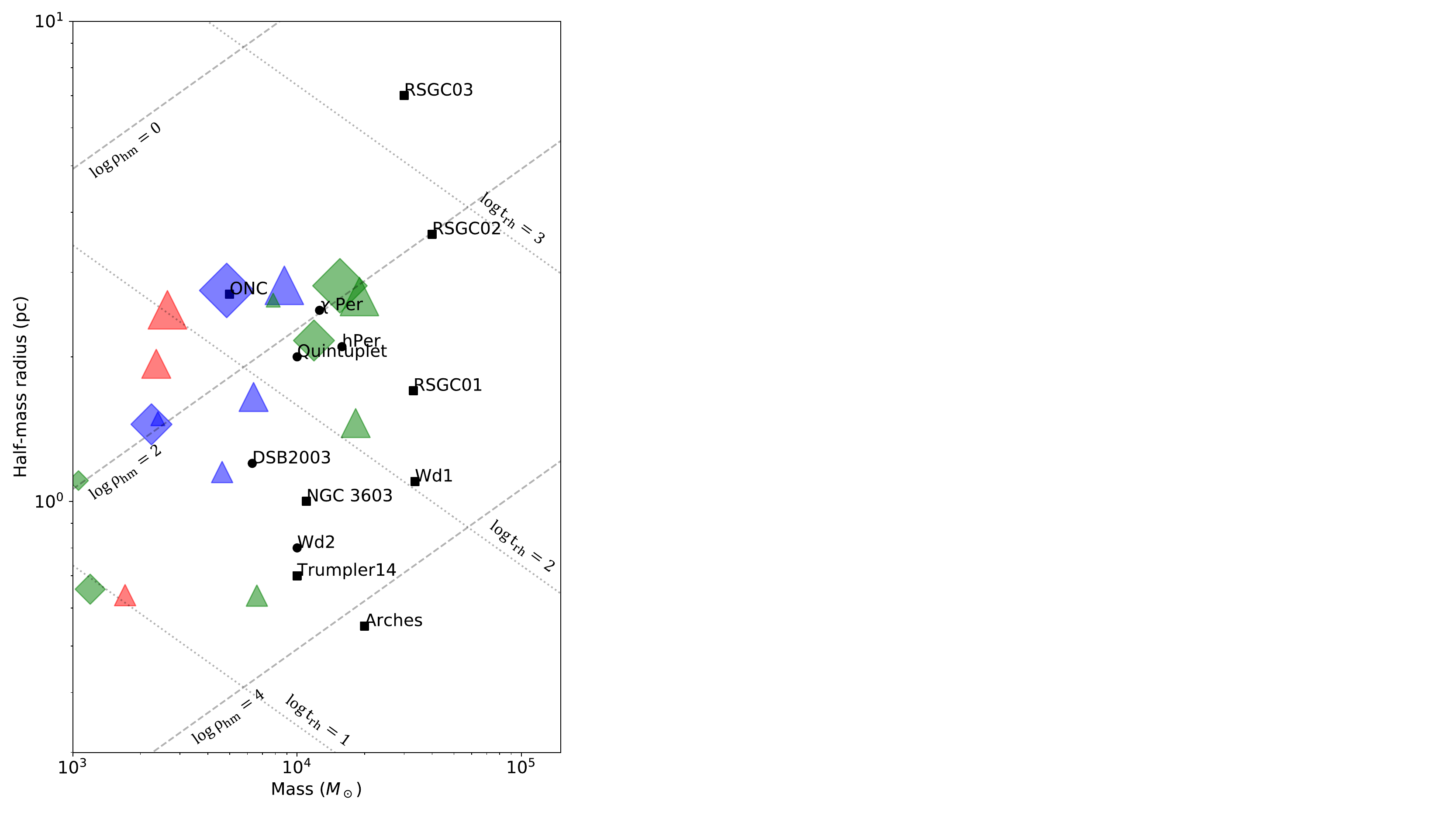}
    \caption{The radius-mass plot extracted from Figure 2 in \protect\cite{zwart_young_2010} that focuses on the young massive clusters is shown. The coloured markers are the clusters in Table \ref{tab:clusters}, whereby the size is directly proportional to the collision speed, the shapes indicate turbulence (triangle: low; diamond: high), and the colours indicate initial cloud density (red: low; blue: standard; green: high). The black squares are the YMCs in the Milky Way from the original plot, whereas the black circle are the additional YMCs in the Milky Way taken from Table 2 in \protect\cite{zwart_young_2010}. Also taken from the original plot are the lines of constant half-mass density $\rho\_{hm}$ and constant half-mass relaxation time $t\_{rh}$.}
    \label{fig:radiusmass}
\end{figure}

Figure \ref{fig:radiusmass} shows the half-mass radius $r\_{hm}$ versus cluster mass $M\_c$ plot for the most massive clusters identified in all models at $t_{10\%}$ and the YMCs in the Milky Way, extracted from Figure 2 in \cite{zwart_young_2010}\footnote{The `mass' and the `half-mass radius' of the real young massive clusters are the photometric mass $M\_{phot}$ and the effective radius $r\_{eff}$.}. Clusters with $M\_c < 10^3$ \msun{} (Clusters H1, H2, H3, H4, L10, H10, and H11) are not shown in Figure \ref{fig:radiusmass} for clarity, but they are located in the region of this plot where open clusters are expected (i.e. $\lesssim 500$ \msun{}). In general, the clusters that are formed with $\mathcal{M}\_{coll} \gtrsim $ Mach 50, $n_0 \gtrsim 236$ \pcm{}, and $\mathcal{M}\_{turb} \approx$ Mach 10 have cluster masses and radii that are the most comparable to the YMCs in the Milky Way, signifying the potential of cloud-cloud collisions to form YMC-like clusters under these extreme initial conditions. Some have similar cluster masses and radii to the YMCs (e.g. Clusters H6 and H8 as compared to the ONC and Quintuplet respectively.) Most of the simulated clusters have a similar $\rho\_{hm}$ of $\sim 10^2$ \msun{}/pc$^3$, while those formed from higher initial cloud density have higher $\rho\_{hm}$. It appears difficult to reproduce denser clusters without adopting a higher initial cloud density. Cluster L12 is the only collision-independent cluster identified by DBSCAN that is in the proximity of YMCs and other clusters formed from high speed and high density collision. Nonetheless, we note that it has a filamentary shape which reflects the high value of $r\_{hm}$. 

YMCs in other high star-forming galaxies usually have mass $\gtrsim 10^4$ \msun{} (e.g. R136 and NGC 1818 in the Large Magellanic Cloud (LMC)) up to $\sim 10^6$ \msun{} (e.g. NGC 1569-A and NGC 1487-1 beyond the Local Group). These YMCs can have radius $\sim$ 10 pc (e.g. NGC 1847 and NGC 1850 in the LMC, and NGC 4449 N-1 beyond the Local Group) \citep[see references listed in Tables 3 and 4,][]{zwart_young_2010}. To achieve such large masses and radii, the precursor clouds would need to have larger cloud mass and sizes compared to our models here, but we make a short comparison of simulations with larger clouds in Appendix \ref{ap:massiveclouds}.


\begin{figure}
    \centering
    \includegraphics[width=\columnwidth,trim={0 0 18.5cm 0}]{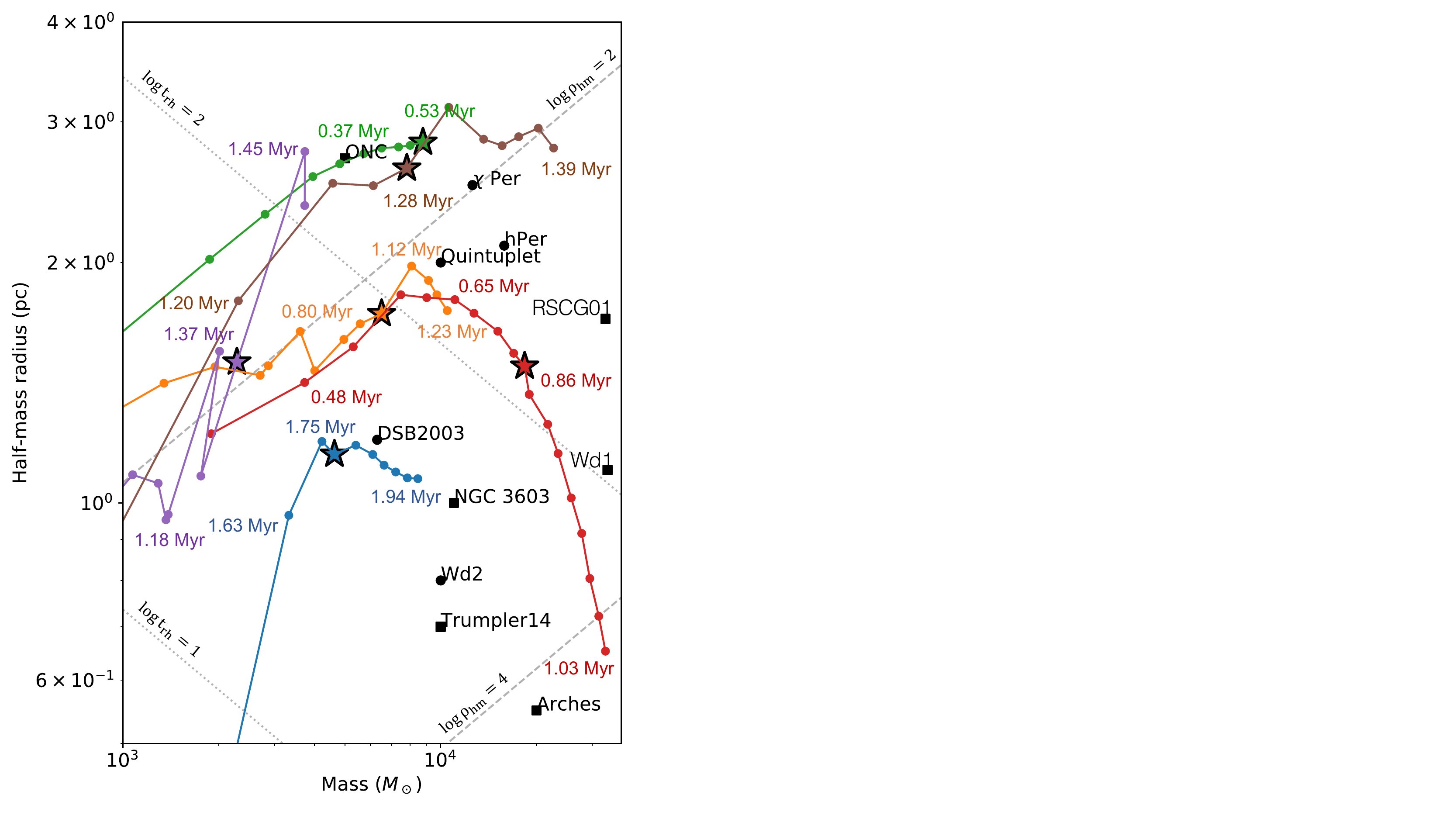}
    \caption{The radius-mass plot similar to Figure \ref{fig:radiusmass} is shown with the evolutionary tracks
    of Clusters L4 (blue), L5 (orange), L6 (green), L8 (red), H5 (purple) and L12 (brown) plotted. The star-shaped markers are the position of the clusters in the radius-mass plot at $t_{10\%}$. The timestamps are coloured according to the cluster colours. As the cluster mass increases over time, the clusters evolve `from left to right' in this radius-mass plot.}
    \label{fig:clusterevolution}
\end{figure}

So far, the properties we list in Tables 2 and 3 are limited to a single time frame. In Figure \ref{fig:clusterevolution} we show how the cluster mass and radius for some of the clusters evolves over time. Cluster L4 (blue), L5 (orange), and L6 (green) are formed via collision of different speeds, Cluster L8 (red) is formed via a collision of higher density clouds, Cluster H5 (purple) is formed via a collision of more turbulent clouds, and finally Cluster L12 (brown) is from a stationary run. The times shown are from the beginning of the simulations. We evolve the simulations for some time after $t_{10\%}$ and track the clusters' evolution. As the clusters accrete more gas and sink particles, and new sink particles are formed in the clusters as they evolve, it is unsurprising that the overall cluster mass increases over time for all clusters. However, the clusters seem to increase in size at earlier times, then decrease later in their evolution. This is because the increase in the number of sink particles at the early phase increases the cluster size, but as the cluster reaches a certain cluster mass, self-gravity becomes significant and thus the cluster starts to relax and contract. This trend is particularly clear for the clusters formed from lower speed collisions with lower turbulence (Clusters L4 and L5; blue and orange), or from collisions with higher density clouds (Cluster L8; red), where the clusters can experience stronger self-gravity compared to other clusters in Figure \ref{fig:clusterevolution}. The same trend of the increasing-then-decreasing radius is likely to happen to other clusters, but either self-gravity is not significant yet and so the distribution of the sink particles is still relatively two-dimensional (Cluster L6; green) or filamentary-like (Cluster L12; brown), or turbulence is too strong which slows down the global gravitational collapse (Cluster H5; purple). We also expect the same trend to occur for other clusters not represented in Figure \ref{fig:clusterevolution}. Figure \ref{fig:clusterevolution} also shows that our colliding models can potentially form clusters with mass and radii similar to some of the YMCs in the Milky Way. For example, Cluster L4 (blue) seems to move towards NGC 3603 in the radius-mass plot, and will likely achieve such mass and radius within another Myr. Cluster L5 (red) can evolve to obtain the mass of Westerlund 1 but about half of its size.

Lastly, we consider the longer term evolution of the clusters by following them in the absence of gas, similar to the treatment by \cite{moeckel_cluster_2010} and \cite{fujii_formation_2015}. Although this is not particularly realistic, it shows what would happen in the event that stellar feedback disperses the remaining gas. We use the pure N-body code \texttt{ph4} in \texttt{AMUSE} \citep{amuse_2018} to simulate the evolution of the cluster under pure gravitational interaction from $t_{10\%}$. We find that, in general, the clusters evolve further towards spherical distributions, during which the clusters lose some members and become smaller. The clusters, especially denser clusters, then remain relatively stable in size and mass for another few Myr and reach the age of the older YMCs observed, such as h-Per and $\chi$-Per of ages $\sim$ 12 Myr.

\section{Summary and discussion}

We have performed simulations of cloud-cloud collisions to study the formation of clusters and in particular YMCs. We investigated how cluster formation depends on the clouds' collision speed, initial cloud density and the level of turbulence. 

The key conclusions are as follows.
\begin{enumerate}

    \item Clusters with greater mass are formed from clouds with higher initial densities, lower turbulence and higher collision speed. The higher collision speed increases the rate of gas convergence into the shock interface and therefore decreases the time to form massive clusters. The morphology of the sink particle distributions (and therefore clusters) also changes with these parameters, so higher collision speed, lower densities and lower turbulence lead to narrower denser shocks, and therefore clusters which reflect the more two-dimensional shape of the shocked region. With lower collision speed, and higher density, the clusters are more spherical as with higher density the clusters experience stronger self-gravity, whilst at lower collision speed the clusters take longer to form, so have longer to undergo gravitational collapse. Higher turbulence decreases the degree of gas convergence onto the shocked region, and reduces the ability of the gas to form massive clusters, but also leads to a more three-dimensional sink particle distribution.
    
    \item The star formation rate increases the fastest for collisions with the highest collision speeds, initial cloud densities and lowest turbulence. Star formation rates can be high in the stationary and low speed models, but this tends to occur at somewhat later times in the simulation and cluster evolution. The difference in star formation rates for the stationary and low speed runs are almost negligible, although the low speed collisions allow the formation of a more massive cluster in the shocked region. We find that for the star formation rate to increase significantly on a shorter timescale, the collision speed has to be at least two times greater than the turbulence, in agreement with results by \cite{matsumoto_2015}.
    
    \item Theoretically, the star formation rate in the shock interface due to collision is proportional to $n_0 \mathcal{M}\_{coll}$, the product of the collision speed and the initial cloud density. Our result shows a positive correlation between the product and the star formation rate, however we find that by incorporating the power-law fit by \cite{krumholz_turbulent_2005} to take into consideration of turbulence and the virial ratio of the clouds, we obtain an improved fit of $\dot{M}_* \propto (n_0 \mathcal{M}\_{coll})^{0.99\pm0.09}$.

    \item We produce clusters which exhibit comparable properties to some of the YMCs in the Milky Way. Our simulations show that cloud-cloud collisions with collision speeds of $\gtrsim$ Mach 50 (relative velocity $\gtrsim 25$ \kms{}), initial cloud density $\gtrsim$ 250 \pcm{}, and turbulence $\sim$ Mach 10 ($\sim 2.5$ \kms{}) lie in a similar space in radius and density, and have formed over timescales of $\lesssim 2$ Myr, in agreement with the short age spreads of YMCs. Lower speed collisions of Mach 20 (relative velocity $\sim$ 10 \kms{}) can form YMC like clusters if the initial cloud density is higher. From Figure \ref{fig:clusterevolution}, we see that increasing the collision speed leads to the formation of clusters on a shorter timescale, but also moves the clusters to larger radii. Similarly, higher turbulence leads to larger cluster radii. Increasing the density of the initial clouds increases the mass of the clusters to higher values. The only simulation of stationary clouds which resulted in a YMC type cluster required the highest (518 cm$^{-3}$) cloud densities, for an already strongly bound cloud. It is not surprising that this cloud produced a massive cluster given the initial densities \citep[see also similar work producing YMCs from single clouds, e.g.][]{fujii_formation_2015,fujii_initial_2015}. However our work shows that it is possible to produce YMCs without requiring such high densities, using densities more comparable with those observed which may be less likely to already be associated with strongly star forming clouds. Using larger size scale clouds would also enable more massive clouds to form, and so it is possible to reduce the initial density required further to produce YMCs \citep{dobbs_liow_2020}. However, we appear to still need fairly high densities in order to produce massive clusters. 

\end{enumerate}

Our simulations highlight various advantages and disadvantages of cloud-cloud collisions as a mechanism for forming YMCs, as compared to, for example, gravitational collapse alone. Firstly, collisions naturally lead to short timescales for forming clusters, compatible with short age spreads. Collisions also focus the production of dense gas and formation of stars to a localised shock region where the clouds collide, even in lower speed collisions. However our simulations show that we still seem to need fairly dense gas to form dense clusters, and clouds which may already be forming stars. Furthermore the morphology of the shocked region leads to clusters which, for the highest speed collisions, are somewhat asymmetric and more cylindrical than spherical in their earliest stages of evolution. Perhaps related to this aspect, our simulations are only able to form very small, dense clusters such as Trumpler 14 if we start with very dense clouds colliding at a moderate relative velocity of $\sim$ 25 \kms{}. It is possible that in a more realistic environment, it is easier to avoid forming aspherical clusters, for example larger scale simulations of colliding galaxies may be more able to form clearly spherical clusters with high velocities gas flows \citep{lahen_merger_2019}.

So far, we only investigate head-on and equal mass cloud-cloud collisions. Collisions between equal mass, more massive clouds, are found to occur in galaxy scale simulations, but are rare  \citep{dobbs_frequency_2015}. An off-axis collision may decrease the overall star formation rate since less gas will be converging at the collision site, whilst off-axis collisions also induce rotation in the clusters \citep{wu_gmc_2015}. Similarly, unequal mass or density cloud-cloud collisions are expected to affect the star formation rate depending on the amount of gas available for star formation at the shock interface. Lastly, unequal size cloud-cloud simulations \citep[e.g.][also performed by us but not shown in this paper]{takahira_formation_2018, sakre_massive_2020} show that the collisions can effectively create arc-like shock interfaces, which then translate into the morphology of the clusters.

We neglect magnetic fields and stellar feedback in our simulations for simplicity.
Magnetohydrodynamical (MHD) cloud-cloud collision simulations by \cite{wu_gmc_II_2017} and \cite{wu_gmc_VII_2019} show that dense filaments are approximately perpendicular to the magnetic field and stronger magnetic field enforces the perpendicularity. A stronger magnetic field also reduces fragmentation which in turn reduces star formation in the clouds. We plan to investigate magnetic fields in future work.

We also do not include stellar feedback, which again we plan to investigate in future work. The timescales of our simulations are quite short, and we would not expect feedback to have a strong impact on the clouds and star formation rate \citep{howard_YMC_2018}. We may expect that feedback stops further star formation as time progresses, although also in these simulations, the clouds have finished colliding with each other on a timescale of a few Myrs, so this would also naturally reduce further star formation. 

\section*{Acknowledgements}

The authors thank Tim Naylor for helpful discussions, and Steven Rieder for the \texttt{AMUSE} N-body code. CLD acknowledges the funding from the European Research Council
for the Horizon 2020 ERC consolidator grant project ICYBOB,
grant number 818940. Simulations for this paper were performed on the DiRAC DIaL service hosted at the University of Leicester, which forms part of the STFC
DiRAC HPC Facility (\url{www.dirac.ac.uk}). The column density plots in this paper were produced using \texttt{SPLASH} \citep{price_splash_2007}. The DBSCAN clustering algorithm was performed using \texttt{scikit-learn} \citep{scikit_learn_2011}.

\section*{Data Availability}

The data underlying this paper will be shared on reasonable request to the corresponding author.

\bibliographystyle{mnras}
\bibliography{mybib} 

\appendix
\section{Resolution Test and Jeans Mass Resolution}
\label{ap:restest}

\begin{figure}
    \centering
    \includegraphics[width=\columnwidth,trim={0 2cm 1.5cm 0}]{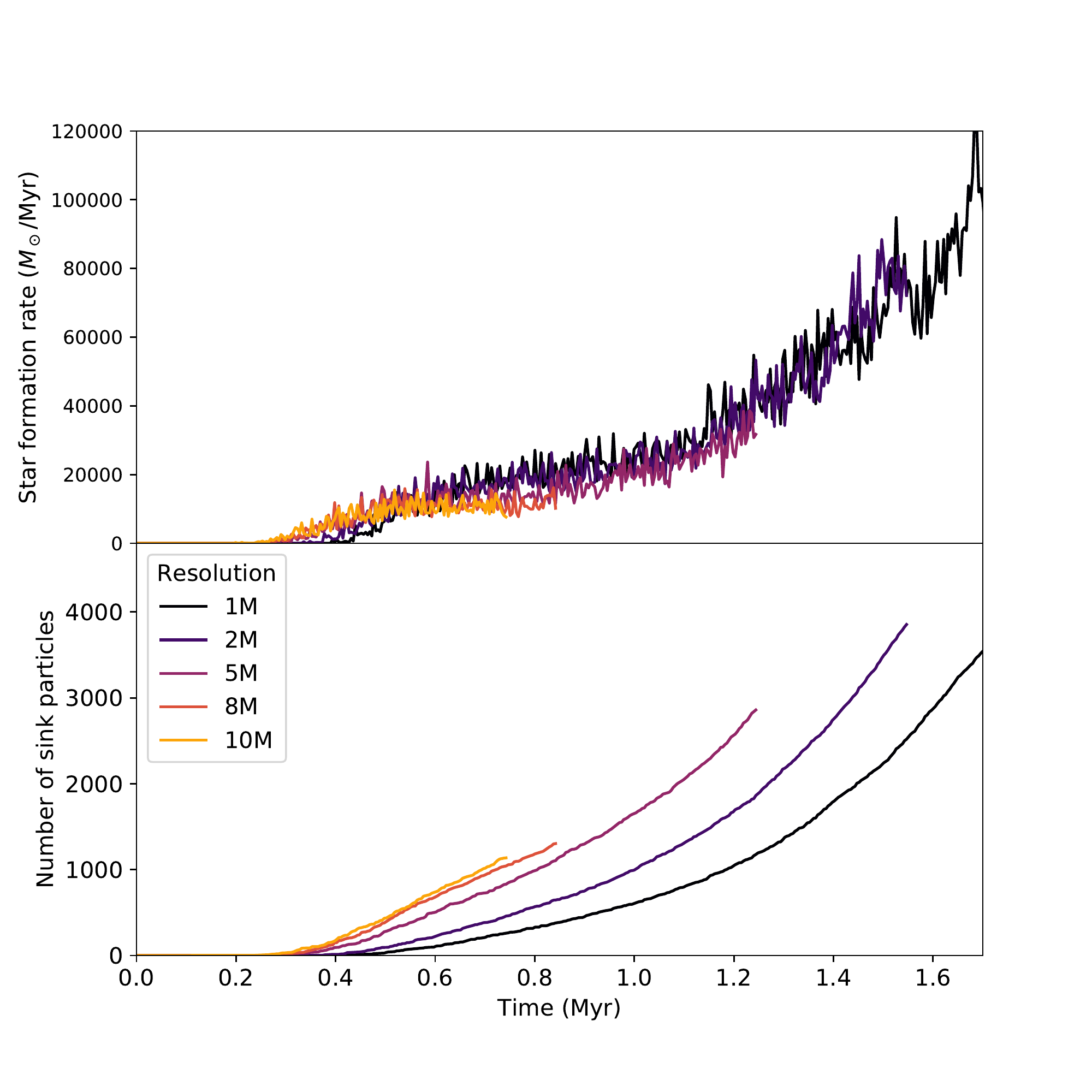}
    \caption{The star formation rate (upper panel) and the number of sink particles formed (lower panel) are plotted against time for the resolution test.} 
    \label{fig:restest}
\end{figure}

We performed a resolution test of a  cloud-cloud collision with $\mathcal{M}\_{coll} = 50$, $n_0 = 236$ \pcm{}, and $\mathcal{M}\_{turb} = 10$ with particle numbers between one and ten million. The result is shown in Figure \ref{fig:restest}. Both the star formation rate and the number of sink particles formed converge as $N\_{SPH}$ increases. We find that $N\_{SPH} = 5 \times 10^6$ is enough to achieve convergence. 

The maximum resolvable density, following the Jeans instability analysis \citep{tohline_hydrodynamic_1982,jeans_2009}, is given as

\begin{equation}
    \rho\_{crit} = \Bigg( \frac{3}{4\pi} \Bigg) \Bigg( \frac{5R\_g T}{2G\mu}\Bigg)^3 \Bigg(\frac{N\_{SPH}}{1.5N\_{neigh}2M\_{cloud}} \Bigg)^2
\end{equation}

\noindent \citep{bate_resolution_1997,bate_formation_2003}, where $R\_g$ is the gas constant. For our simulations and  $N\_{SPH} = 5\times 10^6$, $\rho\_{crit} \sim 10^{-18}$ \gcm{} for the smallest $M\_{cloud}$, and $\rho\_{crit} \sim 10^{-19}$ \gcm{} for the largest $M\_{cloud}$, hence the choice of $\rho\_{sink} = 10^{-18}$ \gcm{} is acceptable. Therefore, our resolution of $N\_{SPH} = 5\times 10^6$ is sufficient to achieve convergence for both the star formation rate and the number of sink particles formed, moreover approximately satisfies the Jeans mass resolution and is computationally efficient. 

\section{Effect of different random turbulent seeds}
\label{ap:turbseeds}

\begin{figure}
    \centering
    \includegraphics[width=\columnwidth,trim={0 0 1.5cm 0}]{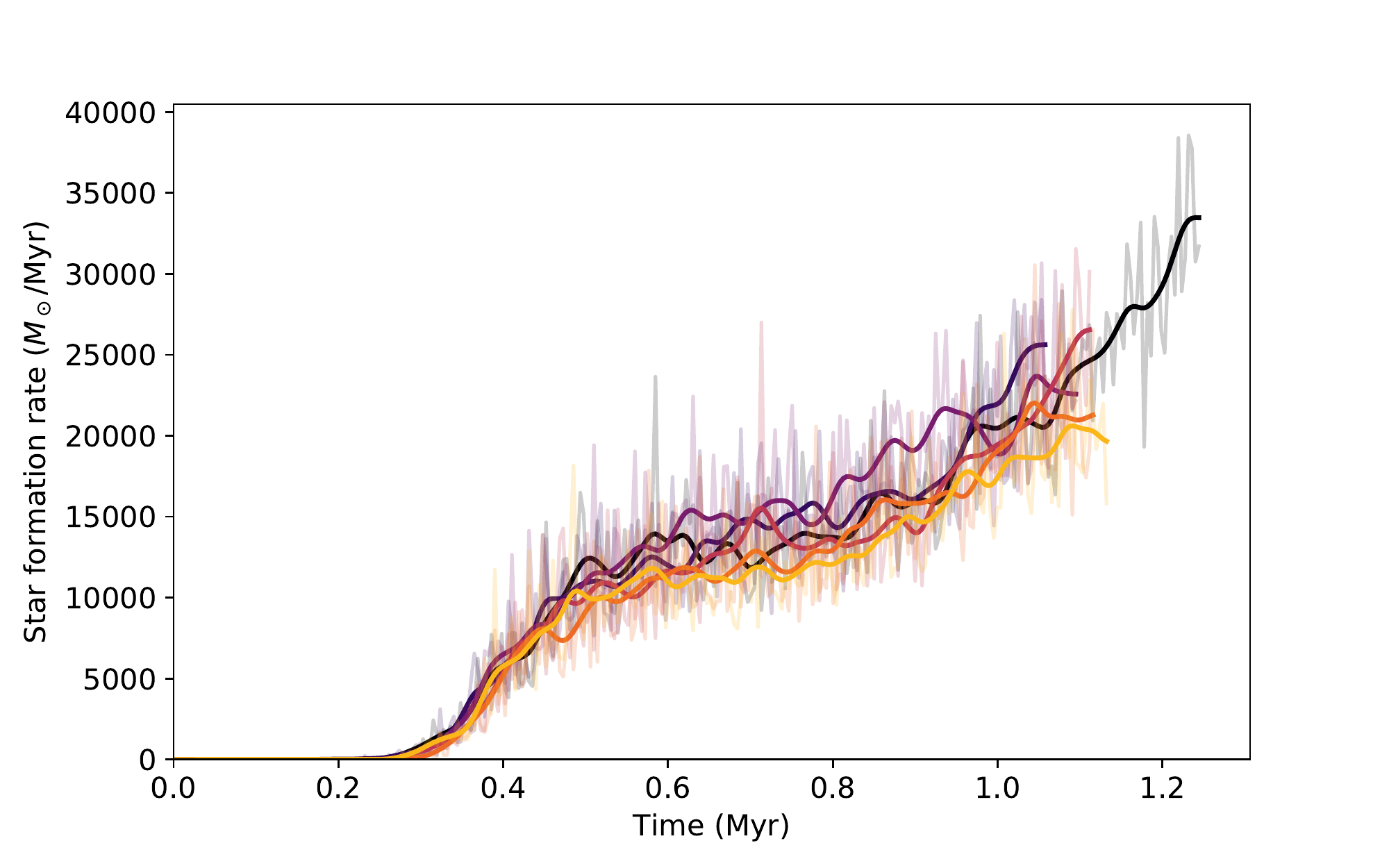}
    \caption{The star formation rates are plotted against time time. All the models here have the same initial conditions, but the random seeds to generate the turbulence fields are different. The black line is the model in the main text, while the rest are the additional five runs.}
    \label{fig:sfrturb}
\end{figure}

We ran five more cloud-cloud collision simulations with standard speed ($\mathcal{M}\_{coll} = 50$), standard density ($n_0=$ 236 \pcm{}) and low turbulence ($\mathcal{M}\_{turb} = 10$), but each with different sets of random seeds than the standard run in the main text to generate the turbulence (initial velocity) field. Figure \ref{fig:sfrturb} shows the star formation rates of these five simulations and the run from the main text. Similar to the method used in Figure \ref{fig:SFR}, a 3$\sigma$-Gaussian filter is used to filter the star formation rates so that the trend can be seen clearly. In general, even though there is some variation in the data, the overall star formation rate of the runs are similar. 

\begin{figure}
    \centering
    \includegraphics[width=\columnwidth,trim={0 0 1.5cm 0}]{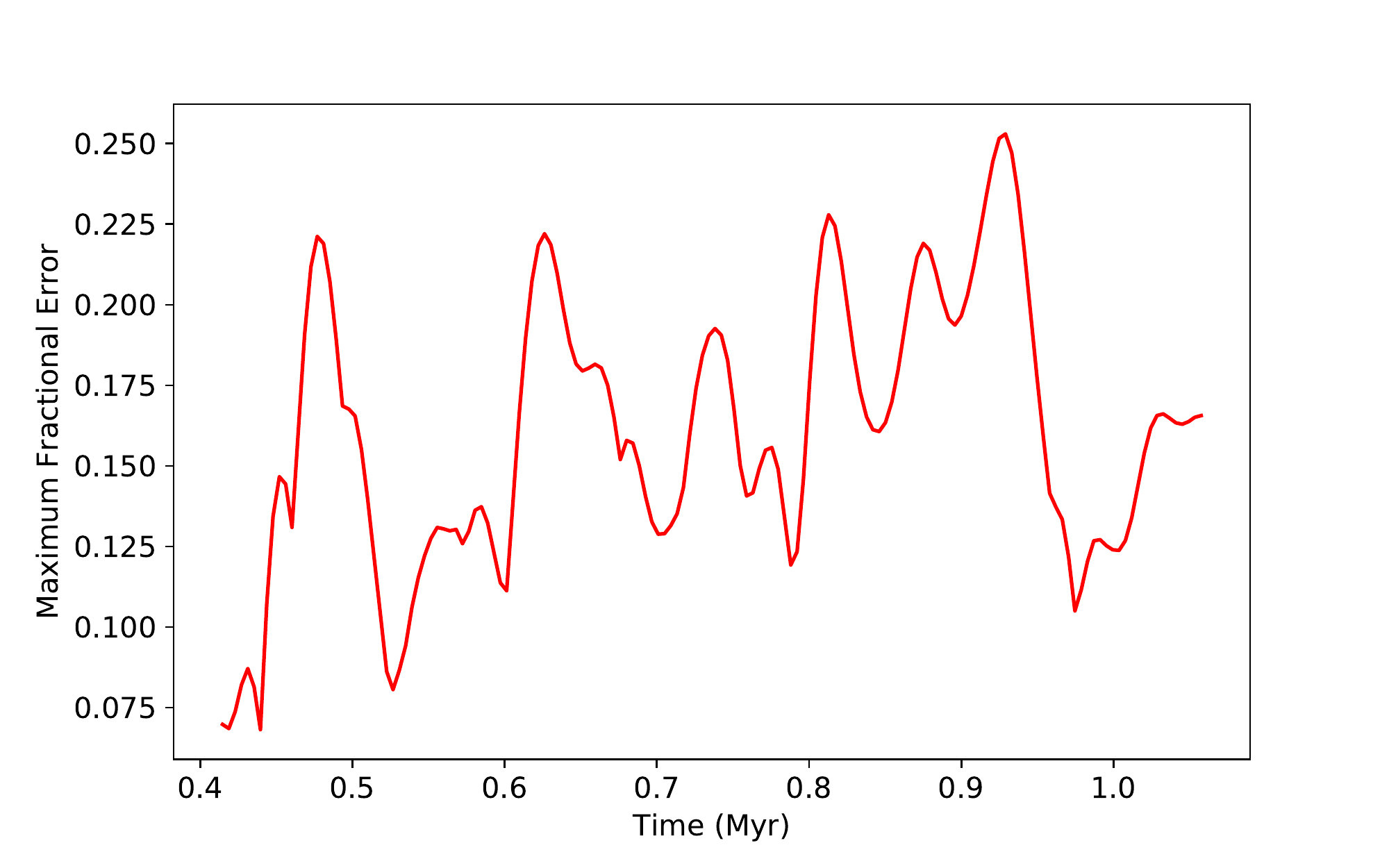}
    \caption{The maximum fractional error from all six runs at any time is plotted at the period when the star formation is active. }
    \label{fig:fracerr}
\end{figure}

Using the filtered star formation rates, we calculate the fractional error of each run with respect to the mean star formation rate, and the maximum of the six fractional errors are plotted against time in Figure \ref{fig:fracerr}. We show the errors at the period when star formation is active, i.e. $t>0.4$ Myr. The average error is $(16\pm4)$\% with maximum error up to 25\%. The result agrees with the variation seen by \cite{dobbs_liow_2020}.

\section{Clusters identified by DBSCAN algorithm}
\label{ap:clusters}

\begin{figure*}
    \centering
    \includegraphics[width=\textwidth,trim = {0 0 13cm 0}]{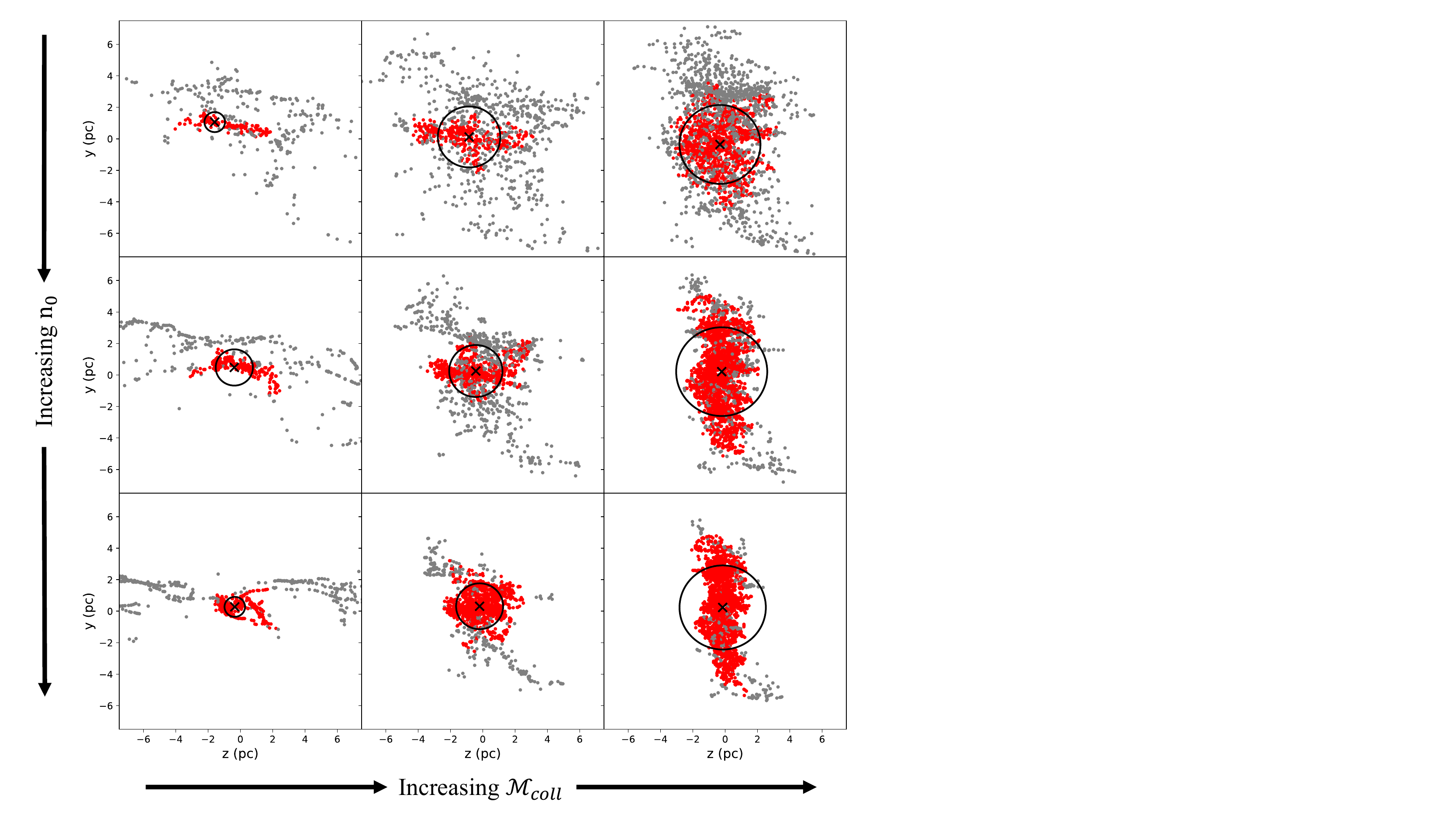}
    \caption{The sink particle distributions in Figure \ref{fig:turb1} are shown with the clusters (red) identified by DBSCAN. The spheres (projected as circles) with radius $r\_{hm}$ and centre-of-mass (cross) enclose the volume with half the cluster mass. The properties of the clusters are presented in the upper section in Table \ref{tab:clusters}.}
    \label{fig:cluster1}
\end{figure*}

\begin{figure*}
    \centering
    \includegraphics[width=\textwidth,trim = {0 0 13cm 0}]{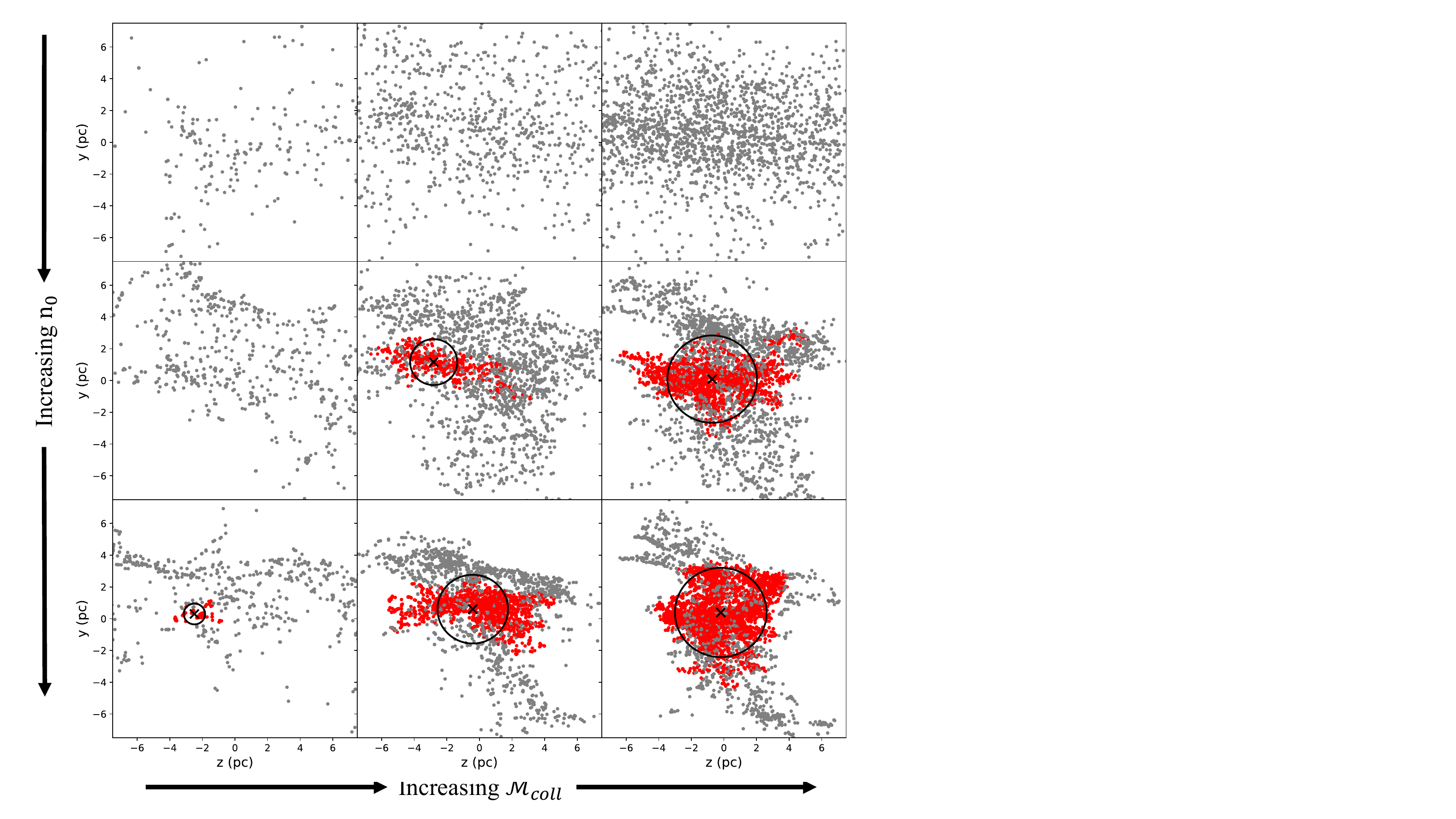}
    \caption{The sink particle distributions in Figure \ref{fig:turb2} are shown with the clusters (red) identified by DBSCAN. The spheres (projected as circles) with radius $r\_{hm}$ and centre-of-mass (cross) enclose the volume with half the cluster mass. The properties of the clusters are presented in the lower section in Table \ref{tab:clusters}.}
    \label{fig:cluster2}
\end{figure*}

Figures \ref{fig:cluster1} and \ref{fig:cluster2} show the most massive clusters identified by DBSCAN algorithm in each model, compared to the underlying distribution of sink particles.

\section{Clusters identified in collisions of larger clouds}
\label{ap:massiveclouds}

Our simulations as described in Section \ref{sec:evolution} follow clouds of $10^4-10^5$  \msun{}, thus they are typical of Milky Way clouds and easier to resolve, but the resultant cluster masses we obtain are naturally limited to the sizes of our initial clouds. We would expect that our results would scale to larger cloud dimensions, but here we present further simulations of collisions of larger clouds to test this. We increase the minor radii and the major radius of our clouds to 9.5 pc and 17.5 pc respectively, and take a mass per cloud of $M\_{cloud} = 10^5$ \msun{}. The cloud density is the same as the "standard density" in the main text, i.e. $\rho_0 = 1.03 \times 10^{-21}$ \gcm{}, or $n_0 = 256$ \pcm{} and it satisfies the Jeans mass resolution criterion. Five simulations with different collision speed and turbulence were performed. Table \ref{tab:clustersmassive} shows the most massive cluster identified by DBSCAN in each of the simulations. Comparing these clusters with the clusters formed in standard density models with smaller clouds (L11, L4$-$L6 and H5), we see that the cluster masses are, as expected, larger for the larger dimension clouds (with the same density, but twice as much mass). We readily form clusters which are $10^4$ \msun{} at an equivalent time. We see, as would be expected, that $r\_{hm}$ is also generally larger in colliding models with larger clouds.

\begin{table*}
    \centering
    \begin{tabular}{c|c|c|c|c|c|c|c|c|c|c|c|c}
    \hline \hline
    Name & $n_0$  & $\mathcal{M}\_{coll}$ & $\mathcal{M}\_{turb}$ & $t_{10\%}$ & $N$  & $z$-COM & $M\_c$          & $r\_{hm}$ & $\rho\_{hm}$      & Age   & $\alpha\_c$ \\
    
        & (\pcm{}) &                      &                      & (Myr)       &      & (pc)    & ($10^3$ \msun{}) & (pc)     & (\msun{}/pc$^{3}$) & (Myr) &            \\
     \hline \hline
     M1 & 236 & 0   & 10 & 1.84 & 539  & -16.4 & 5.80  & 2.29 & 57.5 & 0.52 & 0.69 \\
     M2 & 236 & 20  & 10 & 1.71 & 876  & -0.22 & 12.24 & 0.59 & 6883.7 & 1.12 & 0.49 \\
     M3 & 236 & 50  & 10 & 1.00 & 993  & -0.67 & 9.96  & 1.79 & 203.9 & 0.69 & 0.67 \\
     M4 & 236 & 100 & 10 & 0.52 & 2374 & -0.19 & 15.12 & 3.52 & 41.3 & 0.34 & 0.46 \\
     M5 & 236 & 50  & 20 & 1.27 & 1366 & -0.43 & 9.89  & 2.84 & 51.2 & 0.98 & 0.82 \\

    \hline \hline
    \end{tabular}
    \caption{The list of the most massive clusters formed in the simulations with bigger clouds, with their precursor clouds' initial conditions $\mathcal{M}\_{coll}, \mathcal{M}\_{turb}$ and $n_0$ in the first three columns. The subsequent columns are the time when 10\% of the gas mass becomes sink particles $t_{10\%}$, the number of sink particles $N$, the $z$-component of the cluster's centre-of-mass, the cluster mass $M_c$, the half-mass radius $r\_{hm}$, the half-mass density $\rho\_{hm}$, the cluster age, and the virial ratio $\alpha\_c$. The letter `M' in the names indicates `massive'.}
    \label{tab:clustersmassive}
\end{table*}

\bsp	
\label{lastpage}
\end{document}